\documentclass[fleqn,usenatbib]{mnras}

\usepackage{newtxtext,newtxmath}

\usepackage[T1]{fontenc}
\usepackage{ae,aecompl}

\usepackage{graphicx}
\usepackage{hyperref}
\usepackage{color}
\usepackage[table]{xcolor}
\usepackage{boldline,multirow}
\usepackage{booktabs}
\usepackage{pifont}
\usepackage{mathrsfs}
\usepackage[utf8]{inputenc}

\newcommand{\vmax}{V_{\rm max}}
\newcommand{\rmax}{R_{\rm max}}
\newcommand*\Laplace{\mathop{}\!\mathbin\bigtriangleup}

\usepackage{color}
\definecolor{rosso}{cmyk}{0,1,1,0.4}
\definecolor{rossos}{cmyk}{0,1,1,0.55}
\definecolor{rossoc}{cmyk}{0,1,1,0.2}
\definecolor{blu}{cmyk}{1,1,0,0.3}
\definecolor{blus}{cmyk}{1,1,0,0.6}
\definecolor{bluc}{cmyk}{1,1,0,0.1}
\definecolor{verde}{cmyk}{0.92,0,0.59,0.25}
\definecolor{verdec}{cmyk}{0.92,0,0.59,0.15}
\definecolor{verdes}{cmyk}{0.92,0,0.59,0.4}
\definecolor{bviolet}{rgb}{0.54, 0.17, 0.89}
\hypersetup{colorlinks,bookmarksopen,bookmarksnumbered,citecolor=bviolet,
linkcolor=bviolet,pdfstartview=FitH,urlcolor=bviolet}

\title[Too Big To Fail in Light of Gaia]{Too Big To Fail in Light of Gaia}

\author[]{
Manoj Kaplinghat,$^{1}$ 
Mauro Valli,$^{1}$\thanks{E-mail: mvalli@uci.edu}
Hai-Bo Yu$^{2}$
\\
$^{1}$Department of Physics and Astronomy, University of California, Irvine, CA 92697 USA\\
$^{2}$Department of Physics and Astronomy, University of California, Riverside, CA 92521 USA
}

\date{Accepted XXX. Received YYY; in original form ZZZ}

\pubyear{2019}

\begin{document}
\label{firstpage}
\pagerange{\pageref{firstpage}--\pageref{lastpage}}
\maketitle

\begin{abstract}
We point out an anti-correlation between the central dark matter (DM) densities of the bright Milky Way dwarf spheroidal galaxies (dSphs) and their orbital pericenter distances inferred from Gaia data. The dSphs that have not come close to the Milky Way center (like Fornax, Carina and Sextans) are less dense in DM than those that have come closer (like Draco and Ursa Minor). The same anti-correlation cannot be inferred for the ultra-faint dSphs due to large scatter, while a trend that dSphs with more extended stellar distributions tend to have lower DM densities emerges with ultra-faints. We discuss how these inferences constrain proposed solutions to the Milky Way's too-big-to-fail problem and provide new clues to decipher the nature of DM. 
\end{abstract}




\section{Introduction}
We are in an era where the remarkable success of the standard cosmological model with cold and collisionless dark matter (CDM) \citep{Ade:2015xua} can be tested by observations of dwarf galaxies. The densities probed by dwarfs are sensitive to a wide range of particle physics, including the mass of fermionic DM~\citep{1979PhRvL..42..407T}, DM self-interactions \citep{Spergel:1999mh,Kaplinghat:2015aga}, ultra-light bosonic DM \citep{Hu:2000ke,Hui:2016ltb} and DM as a superfluid ~\citep{Berezhiani:2015bqa}. Milky Way (MW) dwarf spheroidal galaxies (dSphs) are a crucial part of this test \citep{Bullock:2017xww}. Photometry and spectroscopy of these galaxies~\citep{McConnachie:2012vd,Walker:2012td,Battaglia:2013wqa,MNRAS:Simon2019} have led to the formulation of the too-big-to-fail (TBTF) problem~\citep{BoylanKolchin:2011de,BoylanKolchin:2011dk}, based on a key progress in high-resolution simulations of the MW~\citep{Madau:2008fr,Springel:2008cc} and mass estimators of the dSphs~\citep{Walker:2009zp,Wolf:2009tu,Ullio:2016kvy,Errani:2018mnr} tested against hydro simulations \citep{Campbell:2016vkb,Samaniego:2017mnr}.

The dominant interpretation of the TBTF of MW dwarf satellites has been that the CDM N-body subhalos are overly dense. 
This can be alleviated if the MW halo mass is smaller than $10^{12} \rm M_\odot$, but the prevalence of this problem in M31 and in the field \citep{BoylanKolchin:2011dk,Garrison-Kimmel:2014vqa,Papastergis:2014aba} makes this an incomplete solution. 
Another possibility is that the tidal effect of the disk can reduce the inner densities of most of the bright MW satellites~\citep{Brooks:2012vi,Zolotov:2012xd}. 
If satellites have a constant density core in DM, then the impact is even larger~\citep{Penarrubia:2010jk}.
Note that this resolution implicitly assumes dSphs have orbits that take them close to the disk.

Cosmological simulations of the Local Group based on the CDM model, with and without strong feedback effects, have claimed to solve the TBTF problem satisfactorily~\citep{Dutton:2015nvy,Sawala:2015cdf,Wetzel:2016apj}.  
Solutions in the context of non-trivial DM physics including warm DM~\citep{Lovell:2016nkp,Bozek:2018ekc}, DM-radiation couplings ~\citep{Vogelsberger:2015gpr,Schewtschenko:2015rno}, fuzzy DM~\citep{Schive:2014dra} and self-interacting dark matter (SIDM)~\citep{Vogelsberger:2012ku,Zavala:2012us} have also been investigated. The SIDM models that alleviate TBTF require $\sigma/m \gtrsim 1~\rm cm^2/g$~\citep{Zavala:2012us}, which can also explain the DM densities inferred in field galaxies~\citep{Rocha:2012jg,Kamada:2016euw,Ren:2018jpt}. As with CDM, a complete resolution of the TBTF within SIDM (or other models) will require a deeper understanding of environmental effects~\citep{Dooley:2016ajo,Valli:2017ktb}.
 
\begin{figure}
\centering{
\includegraphics[width = \columnwidth]{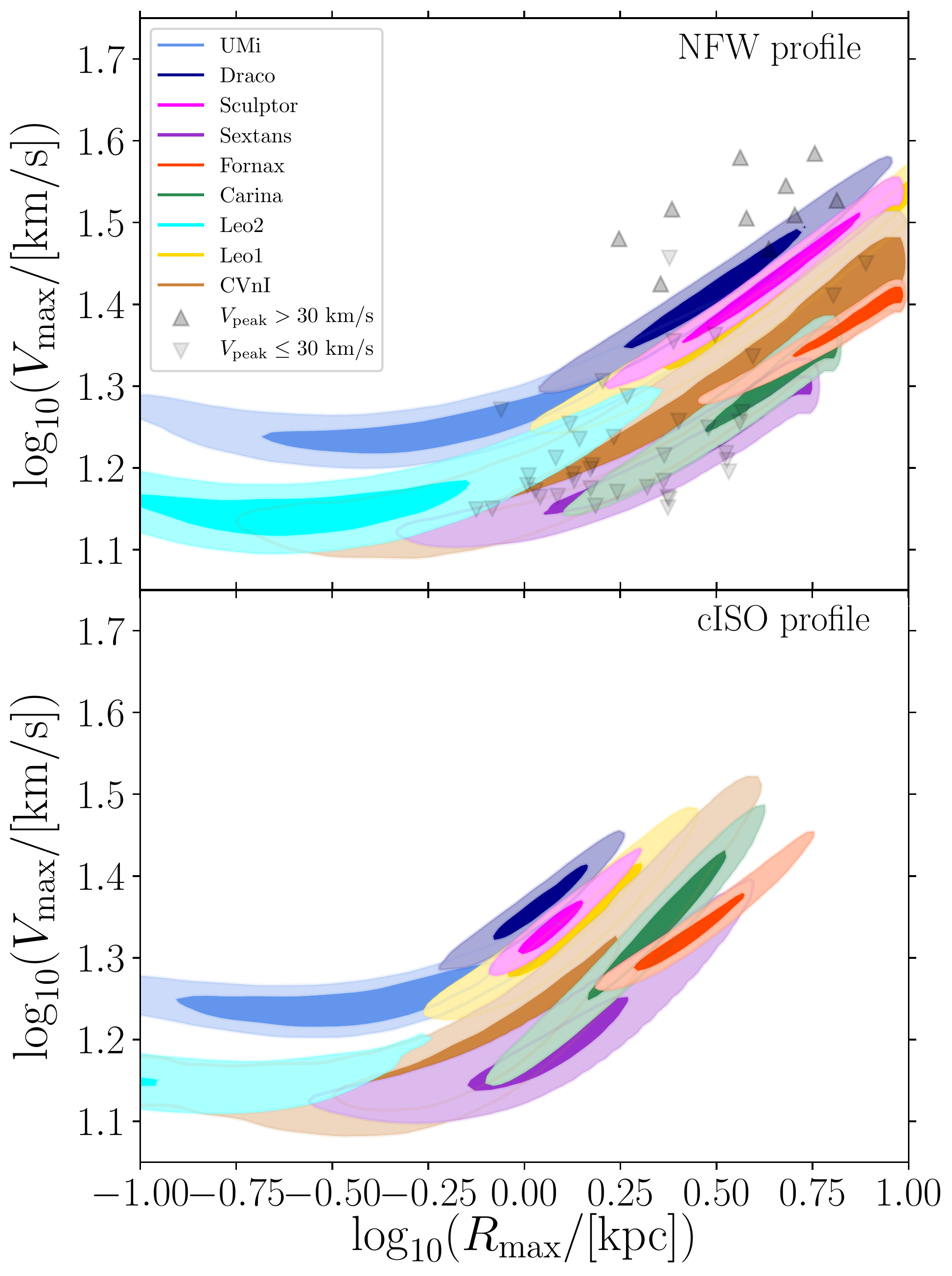}
}
\vspace{-2mm}
\caption{\emph{68\% and 95\% highest probability density in the $\vmax$-$\rmax$ plane for the bright MW dSphs obtained from stellar-kinematics analysis with cuspy and cored DM density profiles. Gray triangle points are 50 subhalos with the highest $\vmax$ over its history ($V_{\rm peak}$) from the CDM ELVIS high-resolution simulation Kauket~\citep{Garrison-Kimmel:2013eoa}. Kauket has the lowest host-halo mass in the suite, $\rm M_{vir} = 1.06\times 10^{12} \rm M_\odot$, but still has subhalos too dense to host the bright dSphs (too-big-to-fail problem). 
}}
\label{fig:MW_subhalo_diversity}
\vspace{-5mm}
\end{figure}

In this \textit{Letter}, we provide a new observational handle on this issue.
We first infer the DM content in the bright MW dSphs through a spherical Jeans analysis~\citep{Battaglia:2013wqa} supplemented with fourth-order projected virial shape parameters~\citep{Aj1990:Merrifield&Kent,Richardson:2014mra,2017MNRAS.471.4541R} and up-to-date photometric information~\citep{MNRAS:Munoz2018}. We find that both cuspy Navarro-Frenk-White (NFW) profile~\citep{Navarro:1996gj} and cored isothermal profile (cISO) can fit the data well and the results are summarized in figure~\ref{fig:MW_subhalo_diversity}. In neither case, do we have to impose cosmological priors to infer the DM densities. For both profiles, we recover the well-known TBTF problem.

We show in figure~\ref{fig:corr_rho_pericenter} that the inferred inner density of the bright MW dwarfs is tightly anti-correlated with the MW pericenter distances~\citep{Fritz:2018aap}, estimated from the most recent observational data provided by the Gaia collaboration~\citep{Gaia:2018dr2a,Gaia:2018dr2b}.
We do not find the same strong anti-correlation for the ultra-faint dSphs; see figure~\ref{fig:corr_rho_pericenter_UDGs}. However, all the dSphs taken together show a clear trend of decreasing DM densities at larger half-light radii in figure~\ref{fig:dSph_size}. 
Our results provide a new perspective for understanding the formation and evolution of the MW dSphs,
as well as a constraint on solutions to the TBTF problem.
We discuss possibilities that could reproduce the observed trends in both CDM and SIDM scenarios.

\section{Jeans analysis and fourth-order virial theorem.}
We adopt the standard spherical Jeans analysis of dSph stellar kinematics~\citep{2008gady.bookB}, and refine it with the inclusion of estimators for fourth-order projected virial theorems~\citep{Aj1990:Merrifield&Kent,Richardson:2014mra}. 
We apply this to the bright dSphs (V-band luminosity $> 10^5 \ \rm L_\odot$ as in~\citet{BoylanKolchin:2011de}), which include the eight classical dSphs and Canes Venatici~I \citep{Zucker:2006he}. The number of available spectroscopic members greatly exceeds a hundred stars for these nine dSphs, which allows for a detailed Jeans analysis.
We analyze the corresponding samples in refs.~\citep{2007ApJ...670..313S,2008ApJ...675..201M,2009AJ....137.3100W,2015MNRAS.448.2717W,2017ApJ...836..202S}, following the methodology in~\citep{Walker:2005nt,Strigari:2007at,Strigari:2008ib,Martinez:2009jh} and in the more recent works~\citep{Pace:2018tin,Petac:2018gue}. 
We perform a Bayesian fit~\citep{ForemanMackey:2012ig,goodman2010} of dSph line-of-sight velocities, 
varying a total of six parameters: the pair of $\vmax$ and $\rmax$, i.e., the maximal circular velocity of the halo and its associated radius, fully determining the DM mass profile, and four parameters to model the stellar orbital anisotropy as in~\citet{Baes:2007tx}.
We describe the stellar density profile by the Plummer model~\citep{1911MNRAS..71..460P} with half-light radii from the recent photometric collection in~\citet{MNRAS:Munoz2018}. 
To impose projected virial theorems, we only use the ``$v_{s1}$" estimator, proportional to $\langle v_{los}^4 \rangle$, developed in refs.~\citep{Richardson:2014mra,2017MNRAS.471.4541R}. 
More details specific to our analysis are provided in appendix~\ref{app:A}.

The mass of MW dSphs is inferred under two separate assumptions for the DM density profile: \textit{i)} cuspy NFW, $\rho = \rho_{s}(r_{s}/r)(1+r/r_{s})^{-2}$, which encapsulates the predictions of N-body CDM simulations, and \textit{ii)} cored isothermal (cISO) profile, which is the prediction of the SIDM model for moderate cross sections when baryons are dynamically irrelevant~\citep{Kaplinghat:2015aga}. 
Both profiles have two free parameters. For the cISO profile, they are 
the central density $\rho_{0}$ and the constant velocity dispersion $\sigma_{0}$. 
These two determine the DM density profile $\rho(r)$ through the Poisson equation: $\sigma^2_0\Laplace\ln\rho=-4\pi G_N\rho$ with the ``cored" boundary conditions at $r=0$: $\rho= \rho_{0}$ and vanishing spatial derivative. Note that both NFW and cISO parameters can be related to ($\vmax$,$\rmax$) pair by: \textit{i)}~$\rho_{s} \simeq 1.721 {\vmax^2}/({G_N\rmax^2}),~r_{s} \simeq 0.462\rmax $; \textit{ii)}~$\rho_{0} \simeq 2.556 {\vmax^2}/({G_N\rmax^2}),~\sigma_{0} \simeq 0.630\vmax $. 

\section{The too-big-to-fail problem revisited}
In figure~\ref{fig:MW_subhalo_diversity}, we show preferred $\vmax$ and $\rmax$ values for the classical MW satellites from our NFW (upper) and cISO (lower) fits, compared to prediction for the 50 most massive subhalos (up-pointing triangles for the most massive ones, i.e. $V_{\rm peak}>30$ km/s, and down-pointing triangles for the others) in the high-resolution ELVIS CDM simulation named Kauket, which has the lowest host-halo mass in the suite~\citep{Garrison-Kimmel:2013eoa}. 
While the choice of the latter goes in the direction of minimizing the TBTF problem, it does not solve it. As figure~\ref{fig:MW_subhalo_diversity} shows, the host halo profiles inferred from both NFW and cISO fits are systematically less massive than those predicted in CDM simulations. It should be noted that both NFW and cISO scenarios provide equally good fits to dSph line-of-sight velocity datasets, while respecting the global constraint from the fourth order projected virial theorem. Details about our fits are in appendices~\ref{app:A}~and~\ref{app:B}, where we also exploit two notions of information criterion~\citep{AkaikeIC,2013arXiv1307.5928G} for model comparison~\citep{BayesFactors}.

The TBTF problem for the cISO profile is particularly interesting in connection with the study in~\citet{Valli:2017ktb}, where
the classical MW dSphs were analyzed using the SIDM halo model of~\citet{Kaplinghat:2013xca,Kaplinghat:2015aga}. The inferred $\sigma/m$ spanned a large range from $0.1\textup{--}0.3~{\rm cm^2/g}$ (Ursa Minor and Draco) to values greater than $10~{\rm cm^2/g}$ (Fornax and Sextans), with a CDM concentration-mass prior from~\citet{Vogelsberger:2015gpr}. This hierarchy of $\sigma/m$ values reflects the diversity of the DM content inferred for the host subhalos, which is not evident in the field halos~\citep{Kamada:2016euw,Ren:2018jpt}. 
Thus, to fully address the MW TBTF problem in CDM or SIDM models, we must understand the physics which is unique to MW dSphs, but is not present for field galaxies. 
As recently investigated in~\citet{Robles:2019mfq}, the interactions of dwarf satellites with the MW disk could play a crucial role in reducing the number of satellites, removing mass from the outskirts of the subhalos and reducing its inner densities both in CDM and SIDM context. Then, it would be natural to expect the dynamical properties of MW dSphs to be correlated with MW pericenter distances, as we discuss next.

\begin{figure}
\centering{
\includegraphics[width = \columnwidth]{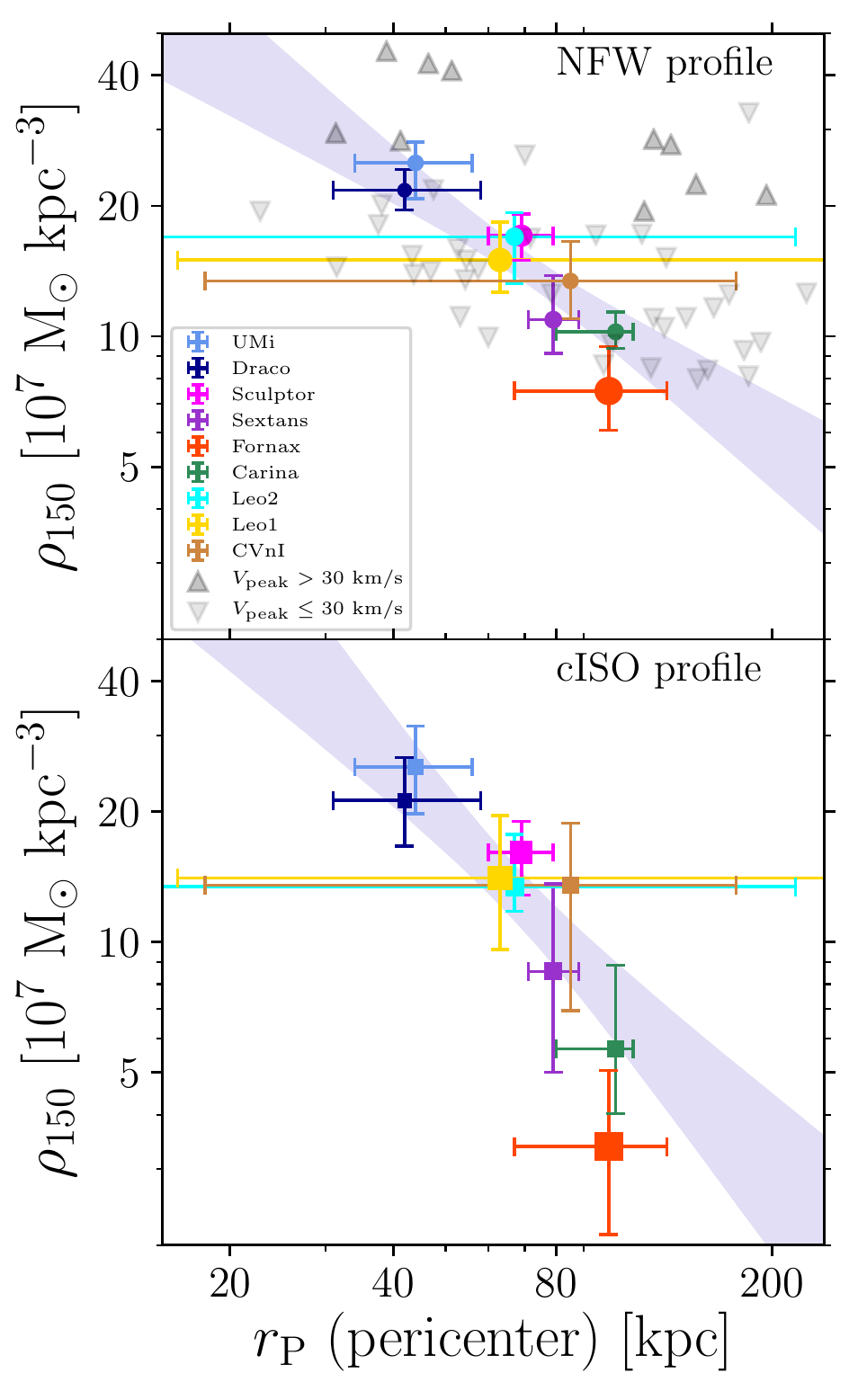}
}
\caption{\emph{
DM density at 150 pc, $\rho_{150}$, inferred from a refined Jeans analysis for the bright MW dSphs vs orbital pericenter distance, $r_{\rm P}$, as estimated in~\citet{Fritz:2018aap} for a MW model with mass of $0.8 \times 10^{12}$~M$_{\odot}$. The upper and lower panels assume that the DM density profile is NFW (cuspy) and isothermal (cored), respectively. In both panels, we show in light violet the 68$\%$ confidence-level region that underlies the correlation between $\rho_{150}$ and $r_{\rm P}$. The size of the points is proportional to the luminosity of the objects.
In the upper panel gray triangles are the densities from the ELVIS subhalos shown in figure~\ref{fig:MW_subhalo_diversity}.}}
\label{fig:corr_rho_pericenter}
\end{figure}

\noindent{\textbf{Pericenters from Gaia: a novel diagnostics.}} The tidal field of the MW could create a correlation between the central density of a satellite and its pericenter passages \citep{1999ApJ...514..109G,2006MNRAS.365.1263M,2006MNRAS.367..387R,Dooley:2016ajo}. In figure~\ref{fig:corr_rho_pericenter}, we show the inferred DM densities at $150$~pc, $\rho_{150}$, from our fits versus the pericenter distances~\citep{Fritz:2018aap}, $r_{\rm P}$, estimated from the recent Gaia data~\citep{Gaia:2018dr2a,Gaia:2018dr2b} for a MW model with mass of $0.8 \times 10^{12}$~M$_{\odot}$. 
For both NFW and cISO fits, there is a tight correlation between the central densities and the pericenter distances of the satellites, in such a way that the denser dSphs have small pericenter distances. 
The correlation persists for a MW model that is twice as heavy or if we adopt density measurements from~\citet{Read:2018pft},while it is no longer present if one trades the pericenter distance with the dSph heliocentric distance, which has been discussed recently, see~\citet{Hammer:2018arx}. These checks and more details on the $\rho_{150}\textup{--}r_{\rm P}$ correlation are available in appendix~\ref{app:C}, where we also show the lack of correlation of $\rho_{150}$ with stellar mass and star formation shutoff time.

The existence of this correlation indicates that the galactic tides play a significant role in shaping the host subhalos of the bright MW bright dSphs, which is to be expected~\citep{Brooks:2012vi}. It is, however, surprising that the central DM density is anti-correlated with the pericenter distance because the tidal effects become more significant as the pericenter distance decreases. One might expect a ``survivor" bias to higher densities in subhalos that venture closer to the MW, but this is not evident in the N-body simulations for the most massive subhalos (upward-pointing triangles in figure~\ref{fig:corr_rho_pericenter}). Adding a disk will preferentially reduce the densities of the smaller $r_{\rm P}$ subhalos and change the orbits. This does not seem to lead to the required correlation~\citep{Robles:2019mfq}, but more investigation is needed to firmly establish this point. 

Note that $150~\rm pc$ is chosen as a compromise between the expectations driven by the dynamics (which constrains the mass within the half-light radius best~\citep{Walker:2009zp,Wolf:2009tu}) and the need for a common scale at where dSph densities can be compared. From figure~\ref{fig:corr_rho_pericenter}, we see that Fornax, Carina and Sextans have lower $\rho_{150}$ values in cISO fits than NFW ones, because their half-light radii are considerably larger than $150$~pc. For the other bright dSphs half-light radii are closer to $150$~pc and consequently $\rho_{150}$ is less sensitive to the choice of the density profile. 

\section{Including ultra-faint dSphs}
A natural question is whether the MW ultra-faint dSphs~\citep{Simon:2007dq,Martin:2007ic,Koposov:2015cua} follow the trend set by the bright dSphs. 
Here we focus on ultra-faint dSphs that have a pericenter distance measurement~\citep{Fritz:2018aap,2018ApJ...863...89S}, $> 2 \sigma$ evidence for non-zero average velocity dispersion, and no proposed association with the Large Magellanic Cloud~\citep{Walker:2016mcs,2018ApJ...867...19K}. From this list, we discard four dSphs: Aquarius~II and Pisces~II due to a small number of spectroscopically-confirmed members; Willman~I due to the large uncertainty on the dynamical-equilibrium condition~\citep{2011AJ....142..128W}; Eridanus~II given the chance that it may not be a MW-bound satellite~\citep{Li:2016utv,Fritz:2018aap}. 
Finally, we also include the recently discovered dSph Antlia~II (AntII), with dispersion and pericenter distance estimates from~\citet{2018arXiv181104082T}.
To infer $\rho_{150}$ conservatively we fit only to the mass estimator of~\citet{Wolf:2009tu}, using measurements and uncertainties tabulated in~\citet{MNRAS:Munoz2018}, except for Reticulum~II, for which we follow~\citet{2018arXiv181012903M}, that includes corrections due to binary systems. Note that Bo{\"o}tes~I  may show some evidence for multiple kinematically distinct populations~\citep{Koposov:2011zi}, not accounted for in the adopted average velocity dispersion.
For the NFW profile, we additionally impose the concentration-mass relation in~\citet{Moline:2016pbm} by adding a statistical weight: $\chi^2_{\rm cosmo} = \big(0.905-\{\log_{10}(c_{200})+[0.146-0.101\,\log_{10}(M_{200}\,10^{-12}/h)]\}/0.3\big)^2$.

\begin{figure}
\centering{
\includegraphics[width = \columnwidth]{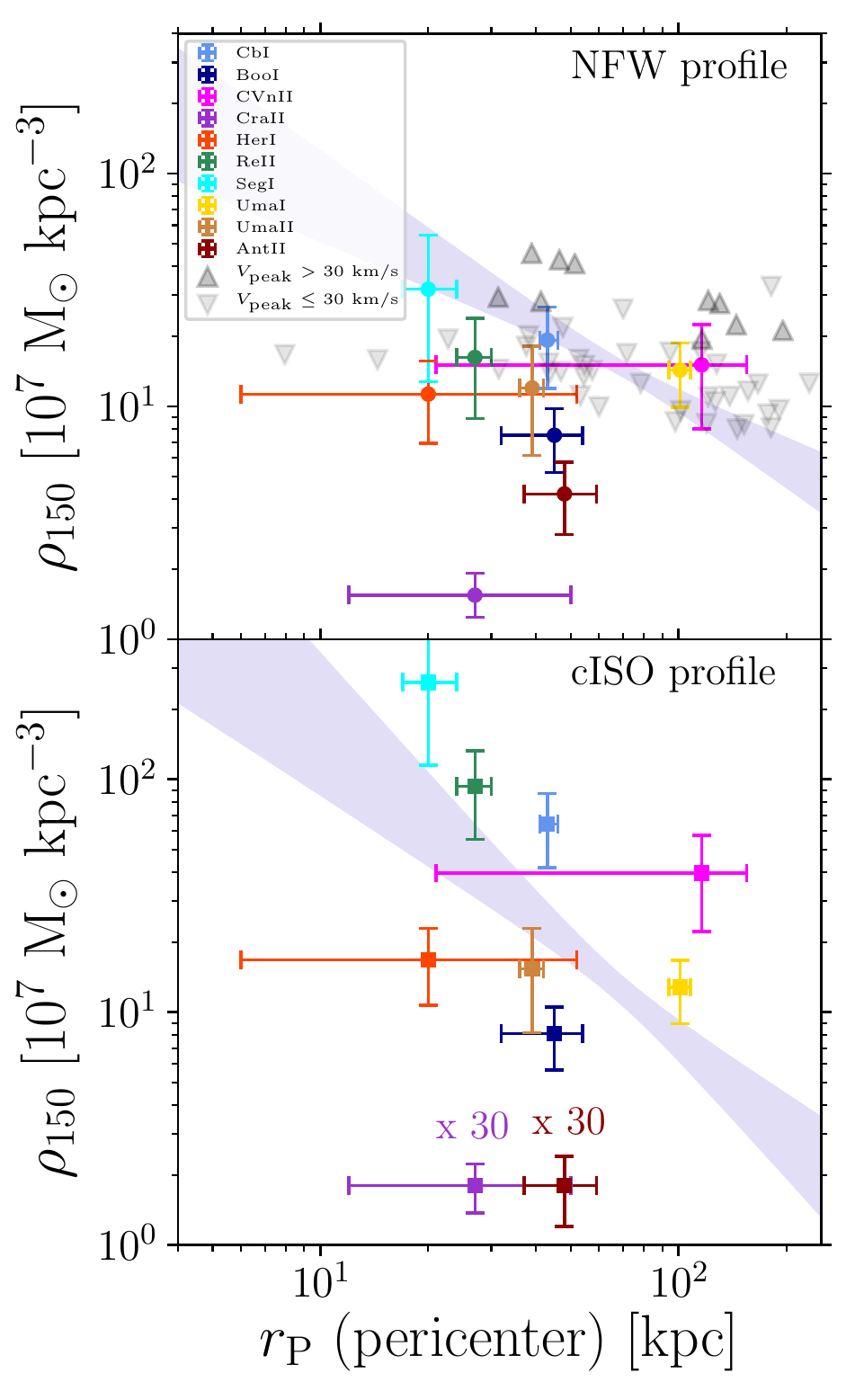}
}
\caption{\emph{ 
The DM density at 150 pc, $\rho_{150}$, versus pericenter distance, $r_{\rm P}$, as in figure~\ref{fig:corr_rho_pericenter} but for ultra-faint dwarfs ($r_{\rm P}$ of AntII is estimated from a MW model with mass of $0.9 \times 10^{12}$~M$_{\odot}$~\citep{2018arXiv181104082T}). 
In both panels, we show the 68$\%$ confidence-level region inferred for the bright dSphs and the gray triangles for the ELVIS subhalos from figure~\ref{fig:corr_rho_pericenter}.}}
\label{fig:corr_rho_pericenter_UDGs}
\end{figure}

Figure~\ref{fig:corr_rho_pericenter_UDGs} shows the inferred $\rho_{150}$ versus pericenter for the selected MW ultra-faint dwarfs. 
Some of them have significantly different $\rho_{150}$ between NFW and cISO fits because the half-light radii quite off from $150~{\rm pc}$. 
Our key result with regard to the ultra-faint sample is the larger scatter in the $\rho_{150}\textup{--}r_{\rm P}$ plane, compared to the anti-correlation for the bright MW dSphs. 
Such a scatter could be due to increased scatter in the stellar-to-halo mass relation at the ultra-faint end and/or reflective of underestimated errors for the ultra-faint dSphs, given their low velocity dispersions and smaller spectroscopic datasets. It may be also connected to DM physics, in particular to DM self-interactions, as discussed later.

The large scatter is evident even without Crater~II (CraII) and AntII, characterized by their extremely low DM density and large stellar extent~\citep{Caldwell:2016hrl,2016MNRAS.459.2370T,2018arXiv181104082T}. This begs the question of why CraII and AntII are so diffuse and underdense (in both DM and stars) compared, for instance, to Draco, whose luminosity and pericenter are similar to those of the two ultra-faint dSphs.

\begin{figure}
\centering{
\includegraphics[width = \columnwidth]{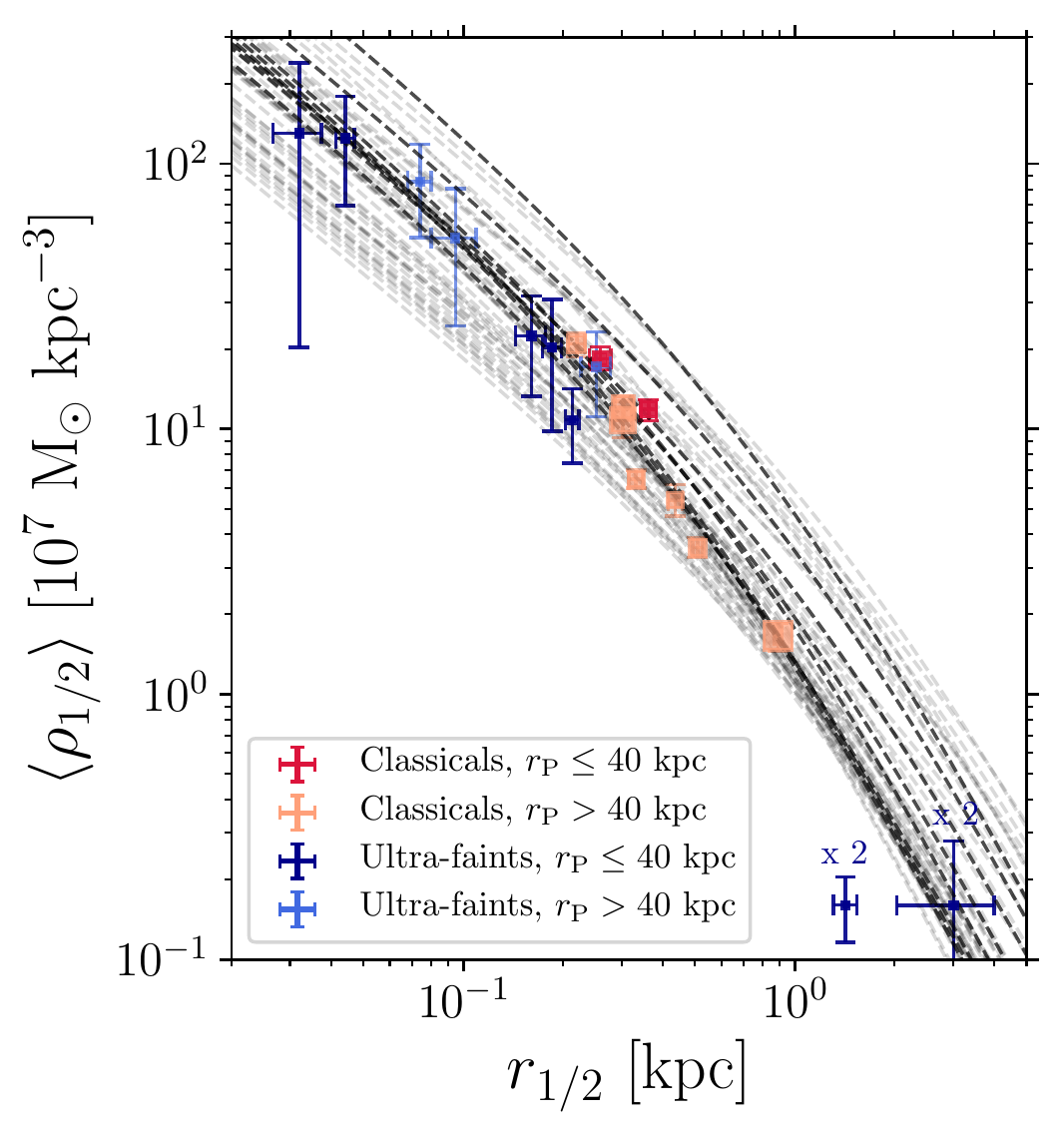}
}
\caption{\emph{Mean density from the mass estimator defined in~\citet{Wolf:2009tu} for all the MW dwarf spheroial satellites considered in the present analysis versus the deprojected stellar half-light radius given in~\citet{MNRAS:Munoz2018}.
 Color code distinguishes nine bright dSphs (red) from ultra-faint dSphs (blue) with estimated pericenter distances smaller (lighter red/blue) or greater (darker red/blue) than $40$~kpc. Darker (lighter) gray lines correspond to mean-density profiles obtained from the same ELVIS subhalos shown in figures~\ref{fig:MW_subhalo_diversity}~-~\ref{fig:corr_rho_pericenter_UDGs}}, with $r_{\rm P}\leq 40$ ($ > 40$) kpc.}
\label{fig:dSph_size}
\end{figure}

To get a global view of the stellar sizes and mass densities of the dSphs, we show in figure~\ref{fig:dSph_size} the mean density at the half-light radius, $\langle \rho_{1/2} \rangle \equiv M_{1/2}/(4/3\pi r_{1/2}^3 )$, versus the stellar half-light radius for all MW dSphs considered in this work.
Note that the inference of $\langle \rho_{1/2} \rangle$ is very robust against the assumed density profile. Consequently, for the bright dSphs we show only NFW fits. For the ultra-faint dSphs we directly derive $\langle \rho_{1/2} \rangle$ from the estimator in~\citet{Wolf:2009tu} using what tabulated in~\citet{MNRAS:Simon2019}.

In figure~\ref{fig:dSph_size} dashed gray lines represent the outcome from CDM ELVIS simulation Kauket~\citep{Garrison-Kimmel:2013eoa}. From the figure one can clearly notice how more extended dSphs favor lower-mass subhalos compared to more compact objects. This interesting trend -- seemingly independent from the pericenter information -- constitutes another unexplained facet of the MW satellites.
The TBTF problem, captured in the upper panel of figure~\ref{fig:MW_subhalo_diversity}, also stands out in the $ \langle \rho_{1/2} \rangle \textup{--}r_{1/2}$ plane of figure~\ref{fig:dSph_size}. In what follows, we discuss the TBTF problem in light of the $\rho_{150}\textup{--}r_{\rm P}$ correlation inferred for the bright MW dSphs.

\section{The TBTF problem in CDM}
To explain TBTF for the bright MW dSphs in light of the Gaia data using collisionless DM models, one must host Draco and Ursa Minor in the densest subhalos with pericenter distances close to 30\textup{--}40 kpc, while explaining why Carina, Fornax and Sextans are hosted by lower-density subhalos that did not come close to the MW disk. 
If the TBTF is solved by reducing the MW mass, then the impact of the disk on the subhalos with smaller $r_{\rm P}$ may render them inconsistent with Draco or Ursa Minor stellar kinematics. Indeed, the effect of a disk in N-body simulations is to reduce the inner densities of the small $r_{\rm P}$ subhalos, but this would leave the other subhalos still too dense compared to data, see e.g.~\citet{Robles:2019mfq}.
Therefore, a full resolution of the TBTF may likely involve physics beyond the tidal impact of the disk.

Feedback from supernovae is one possibility, but its role in setting or diluting the $\rho_{150}-r_{\rm P}$ anti-correlation is unclear. 
Most of the cosmological simulations indicate that below about $10^7~\rm M_\odot$ in stellar mass, feedback is not efficient in changing the density distribution at $1\textup{--}2\%$ of virial radius~\citep{Bullock:2017xww}. 
For subhalo masses around $\rm M_{200} = 10^9~\rm M_\odot$, the relevant radius is about $300~\rm pc$. So it may be still possible for feedback to change the inner density at $150$~pc. However, we have verified that a correlation with $r_{\rm P}$ can still be assessed for densities inferred at $300~\rm pc$. 
In addition, Draco, Ursa Minor, Carina and Sextans have similar V-band luminosities (below $10^6 \ \rm L_\odot$) but they bracket the range of inferred densities.
This would indicate that there is another variable that controls the impact of feedback on DM halos, for example, the ``burstiness" of star formation.
In this regard, we {observe only a weak trend} of our inferred $\rho_{150}$ values with the star formation shutoff times collected in~\citet{Read:2018fxs}, see the appendices.
Interestingly, the difference in the $\rho_{150}$ seen in the sample of dwarfs in~\citet{Read:2018fxs} may reflect the differences in the satellite versus field populations, with the field population having lower $\rho_{150}$.
Clearly, more work along these lines is warranted.

\section{The TBTF problem in SIDM}
SIDM models with cross section over mass $\sigma/m \gtrsim 1 ~\rm cm^2/g$ have been investigated as promising solutions of the TBTF problem~\citep{Vogelsberger:2012ku,Zavala:2012us}. In this regime, a simple model has emerged for the SIDM halo profile~\citep{Kaplinghat:2013xca,Kaplinghat:2015aga,Elbert:2018ApJ,Sameie:2018chj} that explains the diversity of galaxy rotation curves across the full range of galaxy masses~\citep{Kamada:2016euw,Ren:2018jpt,Creasey:2016jaq}. 
However, this halo model does not single out a consistent SIDM cross section when addressing the TBTF problem in the MW~\citep{Valli:2017ktb}.

Recently, \citet{Nishikawa:2019lsc} suggested that for $\sigma/m \gtrsim 5\ \rm cm^2/g$ some of the MW satellites may be in the core-collapse phase, i.e., the central density would be increasing with time~\citep{Balberg:2002ue,Elbert:2014bma}. Interestingly, this process is naturally correlated with how close the satellite comes to the center of the MW. In fact, according to this scenario, Draco would be in the core-collapse phase and hence dense. On the other hand, Fornax, Carina and possibly Sextans would be in the core-expansion phase, similar to the field halos in their evolutionary phase but the central density would be higher due to tidal-mass loss~\citep{Nishikawa:2019lsc}. 
We point out that for a given pericenter, a high-concentration subhalo is more likely to experience core collapse and develop a higher central density than a low-concentration one. This is due to the fact that the core-collapse time scale, $t_c$, depends on concentration, $c_{200}$, as $t_c\propto (\sigma/m)^{-1}M^{-1/3}_{200}c^{-7/2}_{200}$~\citep{Essig:2018pzq,Nishikawa:2019lsc}. For $\sigma/m \simeq 5~\rm cm^2/g$, the relevant time scale can be much larger than the age of the Universe or comparable to it depending on the concentration of the MW subhalo. The low-concentration subhalo, which is not undergoing core collapse, is more vulnerable to tidal effects, further reducing the core density~\citep{Penarrubia:2010jk}. 
For higher concentration subhalos, we expect the density to be larger due to core collapse. This interplay between tidal effects and thermalization of the inner halo due to DM self-interactions is a generic prediction of SIDM models with fairly large cross sections.

\section{Conclusions}
We have fitted stellar kinematics of the nine most luminous MW dSphs and reassessed the TBTF problem for both cupsy and cored DM density profiles. We found a strong anti-correlation between the inner densities of the bright dSphs and their pericenter distances. 
The ultra-faint dSphs show a much larger scatter and the presence of an anti-correlation (if any) is muted.  
These results show that the TBTF problem in light of the Gaia data is still a challenge. We argued that the tidal field of a disk is important but bursty star formation and consequent feedback on DM subhalos may be required in solutions based on CDM models. In SIDM models, MW dSphs could be in either core-expansion and core-collapse phases, depending on their concentrations and pericenter distances. Whether this predicted diversity in the DM content of SIDM subhalos is consistent with data remains to be seen. A more precise knowledge of the central DM densities, orbital motions, stellar sizes and star-formation histories for the MW dSph satellites promises to provide an incisive test of the DM nature.

\section*{Acknowledgements}
We are very grateful to Andrew Pace and Matthew Walker for their help with the data and useful discussions. We thank Mike Boylan-Kolchin, James Bullock, Shea Garrison-Kimmel and Joshua Simon for interesting comments. 
M.K. and M.V. are supported by the NSF Grant No.~PHY-1620638. H.B.Y. acknowledges support from U.S. Department of Energy under Grant No.~de-sc0008541 and UCR Regents' Faculty Development Award.




\bibliographystyle{mnras}
\bibliography{mnras_TBTF_Gaia} 



\appendix
\onecolumn

\section{Jeans analysis refined by fourth-order virial shape parameters}
\label{app:A}
The spherical Jeans analysis has been extensively used in literature to analyze the dynamics in MW dSphs via the measured stellar kinematics; see refs.~\citep{Bonnivard:2015xpq,Petac:2018gue} for recent analyses reviewing in detail this approach. 
Under spherical symmetry approximation,\footnote{{This is well-motivated by MW dSph ellipticities inferred from  photometric data, see for instance~\cite{MNRAS:Munoz2018}. 
Small systematics related to departure from spherical symmetry have been assessed via numerical simulations~\citep{Bonnivard:2014kza,Campbell:2016vkb,Samaniego:2017mnr,Errani:2018mnr}, and Schwarzschild modeling~\citep{2013ApJ...763...91J,2013A&A...558A..35B,2017MNRAS.470.3959K,2018MNRAS.476.2918K,2019MNRAS.482.5241K}.  
Axi-symmetric Jeans analysis~\citep{Cappellari:2008kd} has been also performed under simple specific ansatz on the stellar orbital anisotropy~\citep{Hayashi:2012,Hayashi:2016kcy}.}}
 Jeans equations~\citep{2008gady.bookB} reduce to:
\begin{eqnarray}
\label{eq:Jeans_eq}
\left(\frac{d\log\left(\nu \sigma_{r}^2 \right)}{d \log r}  + 2   \beta \right)\sigma_{r}^2 = - G_{N} \frac{M}{r} \ .
\end{eqnarray}
with $\nu$ being the stellar density profile of the system, $\beta$ the stellar velocity dispersion anisotropy and 
$M$ is the total mass profile, which is well-approximated by the DM component for dSphs~\citep{Walker:2012td,Battaglia:2013wqa}. 
The parameterization of $\beta$ adopted from~\citet{Baes:2007tx} is: 
\begin{eqnarray}
\label{eq:beta}
\beta(r) =\frac{\beta_{0} + \beta_{\infty} (r/r_{\beta})^{\eta}}{1 + (r/r_{\beta})^{\eta}} \ .
\end{eqnarray}
Solving equation~\eqref{eq:Jeans_eq} for $\sigma_{r}^2$, one can make contact with observations through the line-of-sight projection:
\begin{eqnarray}
\label{eq:sigma_los}
\Sigma \, \sigma_{los}^{2} = \int_{R^2}^{\infty} \frac{d r^2}{\sqrt{r^2-R^2}} \left(1-\beta \frac{R^2}{r^2} \right) \nu \sigma_{r}^2 \ .
\end{eqnarray}
In the above, $\Sigma$ is the surface brightness of the system, related to $\nu$ via an Abel transform~\citep{Wolf:2009tu,Ullio:2016kvy} 
and constrained by available photometric data, while $\sigma_{los}$ is the line-of-sight velocity dispersion profile of stars inferred from spectroscopy of individual stars in the dSph. 
In this work, we model the surface brightness with a Plummer model, i.e. $\Sigma \propto (1+R/R_{1/2})^{-2}$. The structural parameters of the Plummer model, projected half-light radii $R_{1/2}$ and ellipticities, are taken from the recent work of~\citet{MNRAS:Munoz2018} (see also \citet{MNRAS:Simon2019}). We modify the semi-major half-light radius by a factor of $\sqrt{1-\epsilon}$ to be consistent with our assumption of spherical symmetry.
{In the present Jeans analysis for the brightest MW dSphs we also include the contribution of the stars to the potential well of the system for the objects where such an effect turns out to be more important than a few percent level for the inference of $\rho_{150}$. This is actually the case for Fornax, Leo~I and Sculptor, for which we adopt the total stellar mass reported in~\cite{Read:2018fxs}, describing the stellar contribution to the internal dynamics of the system with the Plummer model.}
 
The sample of line-of-sight velocities for the nine bright MW dSphs is publicly available from refs.~\citep{2007ApJ...670..313S,2008ApJ...675..201M,2009AJ....137.3100W,2015MNRAS.448.2717W,2017ApJ...836..202S} (for Ursa Minor data, M.G.~Walker, private communication) and using stellar-membership cuts obtained in~\citet{Pace:2018tin} (A.Pace, private communication). We adopt dSph RA-Dec coordinates and heliocentric distance from~\citet{MNRAS:Munoz2018} and infer the average bulk velocity $\overline{V}$ and dispersion $\overline{\sigma}_{los}$ using the Gaussian estimator proposed in~\citet{Walker:2005nt}. We then subtract the estimated $\overline{V}$ from the line-of-sight velocities 
in order to remove the bulk motion of the dSph.

The test statistic to model the probability of obtaining the dataset $\{v_{los}\}_{i=1,\dots,N_{\star}}$ given the model parameters involved in our analysis corresponds here to the Gaussian likelihood~\citep{Strigari:2007at,Martinez:2009jh}:
\begin{eqnarray}
\label{eq:Lvlos}
 \mathcal{L}_{v_{los}}  =  \prod_{i=1}^{N_{\star}} \frac{1}{\sqrt{2 \pi \left(\delta v_{los,i}^2 + \sigma_{los}^2(R_{i})\right)}} \exp\left(-\frac{1}{2} \frac{v_{los,i}^2}{\delta v_{los,i}^2+\sigma_{los}^2(R_{i})}\right) \ ,
\end{eqnarray}
where 
$\delta v_{los,i}$ and $R_{i}$ are the velocity measurement error and position of the $i$-th star, 
and $\sigma_{los}^2(R_{i})$ corresponds to the intrinsic dispersion predicted by equation~\eqref{eq:sigma_los}. 
We use the public package \textit{emcee}~\citep{ForemanMackey:2012ig} in order to perform Monte Carlo Markov Chain (MCMC) sampling. Our theoretical model comprises of 6 parameters: the four characterizing $\beta$ in equation~\eqref{eq:beta}, and ($\vmax$, $\rmax$) pair determining the DM halo profiles considered in this work. For the stellar anisotropy $\beta$, we adopt the priors of~\citet{Valli:2017ktb}, which take into account the conditions stemming from the requirement of non-negative phase-space distribution functions.
For the DM halo, we impose the broad priors: $-2 \leq \log_{10}(\vmax/[\textrm{km}/\textrm{s}])\leq 2$ and $-2 \leq \log_{10}(\rmax/\textrm{kpc})\leq \log_{10}(r_{\textrm{J}}/\textrm{kpc})$. We demand that $\rmax$ be smaller than the instantaneous Jacobi radius $r_{\textrm{J}}$, which is a reasonable description of $\rmax$ for subhalos in CDM simulations \citep{Springel:2008cc}. This restriction is mostly relevant for the NFW profile and it removes some of the large $\rmax$ solutions, which is evident in the contour shape at high $\rmax$ in figure~\ref{fig:MW_subhalo_diversity}. 

We set the Jacobi radius $r_{\textrm{J}} \equiv (G_{N}M_{\textrm{dSph}}D^2/(2 \sigma_{\textrm{MW}}^2))^{1/3}$ similarly to what was done previously (e.g., ref~\citep{Strigari:2008ib}). 
We approximate the distance $D$ of the satellite from the MW by the dSph heliocentric distance, and take 200 km/s as an estimate 
for the MW velocity dispersion $\sigma_{\textrm{MW}}$. Since the restriction on $\rmax$ does not have a significant impact on the physics discussion, we use estimates of the total mass of the dSphs $M_{\textrm{dSph}}$ from~\citet{Errani:2018mnr} for the cored and cuspy cases.
Finally, we note that for the cISO fit of CVn~I, a second mode is found at extremely low $\rmax$ values: we have cut this second solution restricting for this case the prior on $\rmax$ to values $\log_{10} \rmax\geq-1$.

We refine the standard Jeans analysis with the constraints from the fourth-order projected virial theorems~\citep{Aj1990:Merrifield&Kent} for approximately spherical systems~\citep{Richardson:2014mra}:
\begin{eqnarray}
\label{eq:fourth_order}
v_{1s} \equiv \frac{1}{2} \int_{0}^{\infty} dR^2 \, \Sigma  \, \langle v_{los}^4 \rangle   & = & \,  \frac{G_{N}}{5} \int_{0}^{\infty} d r^2 \,  M \left(5 - 2 \beta \right) \nu \sigma_{r}^2 \ , \\
v_{2s} \equiv \frac{1}{2} \int_{0}^{\infty} dR^2 R^{2} \, \Sigma \, \langle v_{los}^4 \rangle & = & \,  \frac{2 G_{N}}{35} \int_{0}^{\infty} d r^2 r^2 \,  M \left(7 - 6 \beta \right) \nu \sigma_{r}^2  \ , \nonumber 
\end{eqnarray}
where $\langle v_{los}^{4}\rangle$ denotes the fourth moment of the line-of-sight velocity distribution. The left-hand side of equation~\eqref{eq:fourth_order} involves only observational quantities and is determined by the following means 
$\{v_{los,i}^4\}_{i,\dots,N_{\star}}$ and $ \{R_{i}^2v_{los,i}^4\}_{i,\dots,N_{\star}}$~\citep{2017MNRAS.471.4541R}.

We compute mean and variance for the estimators of $v_{1s}$ and $v_{2s}$ by generating $10^6$ sets of the samples $\{v_{los,i}^4\}_{i,\dots,N_{\star}}$ and $ \{R_{i}^2 \, v_{los,i}^4\}_{i,\dots,N_{\star}}$ via the distribution encoded in equation~\eqref{eq:Lvlos} for $v_{los}$, namely a Gaussian with mean 0 and variance approximated by $\overline{\sigma}_{los}^2$. Another possibility suggested by~\citet{2017MNRAS.471.4541R} would be to directly compute the surface brightness integral in equation~\eqref{eq:Lvlos} within a given functional parameterization for $\Sigma$ and a data-based interpolation for the radial profile of $\langle v_{los}^4 \rangle$. 
We tested this alternative procedure and found the estimate of $v_{1s}$ to be quite stable against the method adopted. The estimate of $v_{2s}$, instead, turns out to be strongly sensitive to the assumed outer behavior for $\Sigma$ and $\langle v_{los}^4 \rangle$ profiles at large radii (e.g.: for a Plummer surface brightness profile and a mildly varying $\langle v_{los}^4 \rangle$ profile, $\nu_{2s}$ would formally diverge), implying there is a large uncertainty on the corresponding data-based estimator. Therefore, we only use constraints from $v_{1s}$:
\begin{eqnarray}
\label{eq:VSP}
  \widehat{v_{s1}} \equiv \frac{1}{2\pi N_{\star}}\sum_{i=1}^{N_{\star}} v_{los,i}^{4}  \, \simeq \,   \frac{G_{N}}{5} \int_{0}^{\infty} d r^2 \,  M \left(5 - 2 \beta \right) \nu \sigma_{r}^2 \ .
\end{eqnarray}
From our Monte Carlo sampling of $v_{los}$, the data-based estimator $\widehat{v_{s1}}$ can be reasonably approximated by a Gaussian distribution. We report the 16-$th$, 50-$th$ and 84-$th$ percentiles of the inferred distribution in table~\ref{tab:results}. We then use the estimated mean and variance for $\widehat{v_{s1}}$ to build a Gaussian weight, $\mathcal{L}_{v_{s1}}$, in our fit that supplements the observational information from the likelihood in equation~\eqref{eq:Lvlos} and refines our Jeans analysis according to equation~\eqref{eq:VSP}.

In the same table~\ref{tab:results} we show the results from our Bayesian inference of $(\vmax,\rmax)$ pair for both cuspy and cored DM profiles considered in this work, reporting again 16-$th$, 50-$th$ and 84-$th$ percentiles of the related posterior distribution. For a rough estimate of the goodness of the fit, we compute $\chi^2/\textrm{d.o.f.} \equiv -2\log \widetilde{\mathcal{L}}_{v_{los}}/(N_{\star}-(6-1))$, where $\widetilde{\mathcal{L}}_{v_{los}}$ corresponds to the product of the Gaussian distributions in equation~\eqref{eq:Lvlos} without the normalization factor.
As shown in table~\ref{tab:results}, the constraint from $\eqref{eq:fourth_order}$ does not forbid an overall good fit to dSph stellar kinematics in any of the cases considered. To compare the cored and cuspy density profile fits, we estimate the information criteria: 
(i) the frequentist Akaike Information Criterion (AIC)~\citep{AkaikeIC}, related to the log-likelihood $\log \mathcal{L}_{v_{los}+v_{1s}}$ at the global mode of the fit, and (ii) the Bayesian deviance Information Criterion (BIC)~\citep{2013arXiv1307.5928G}, obtained from the mean and variance of the posterior distribution of $\log \mathcal{L}_{v_{los}+v_{1s}}$. Note that, differently from the$\chi^2/\textrm{d.o.f.}$ defined above, both AIC and BIC are sensitive to the goodness of fit to the line-of-sight velocity dispersion and the global constraint from the estimated fourth-order virial-shape parameter $\nu_{s1}$. 
For both AIC and BIC, the standard scale of evidence reported e.g. in~\citet{BayesFactors} suggests that differences (in absolute value) of $\mathcal{O}(10)$ or greater in the information criteria point to data-based preference for the model with the lower value of AIC or BIC.

Interestingly, we observe that $\Delta\textrm{AIC} \equiv \textrm{AIC}_{\rm NFW}-\textrm{AIC}_{\rm cISO}$ seems to highlight some mild preference for the cored halos in Sculptor and Fornax. We verified that this would no longer be the case if we would remove the information stemming from $v_{s1}$ in the corresponding fits.
At the same time, $\Delta \textrm{BIC} \equiv \textrm{BIC}_{\rm NFW}-\textrm{BIC}_{\rm cISO}$, while confirming the trend highlighted by the AIC, never exceeds $\mathcal{O}(1)$ values, pointing to the fact that the assumed cored and cuspy halo profiles provide roughly the same level of goodness in the description of dSph stellar kinematics. We conclude that the inclusion of the information from $\nu_{s1}$ in the spherical Jeans analysis does not effectively break the well-known $M$~-~$\beta$ degeneracy for MW dSphs. 

\begin{figure}
\vspace{10mm}
\centering{
\includegraphics[width = \columnwidth]{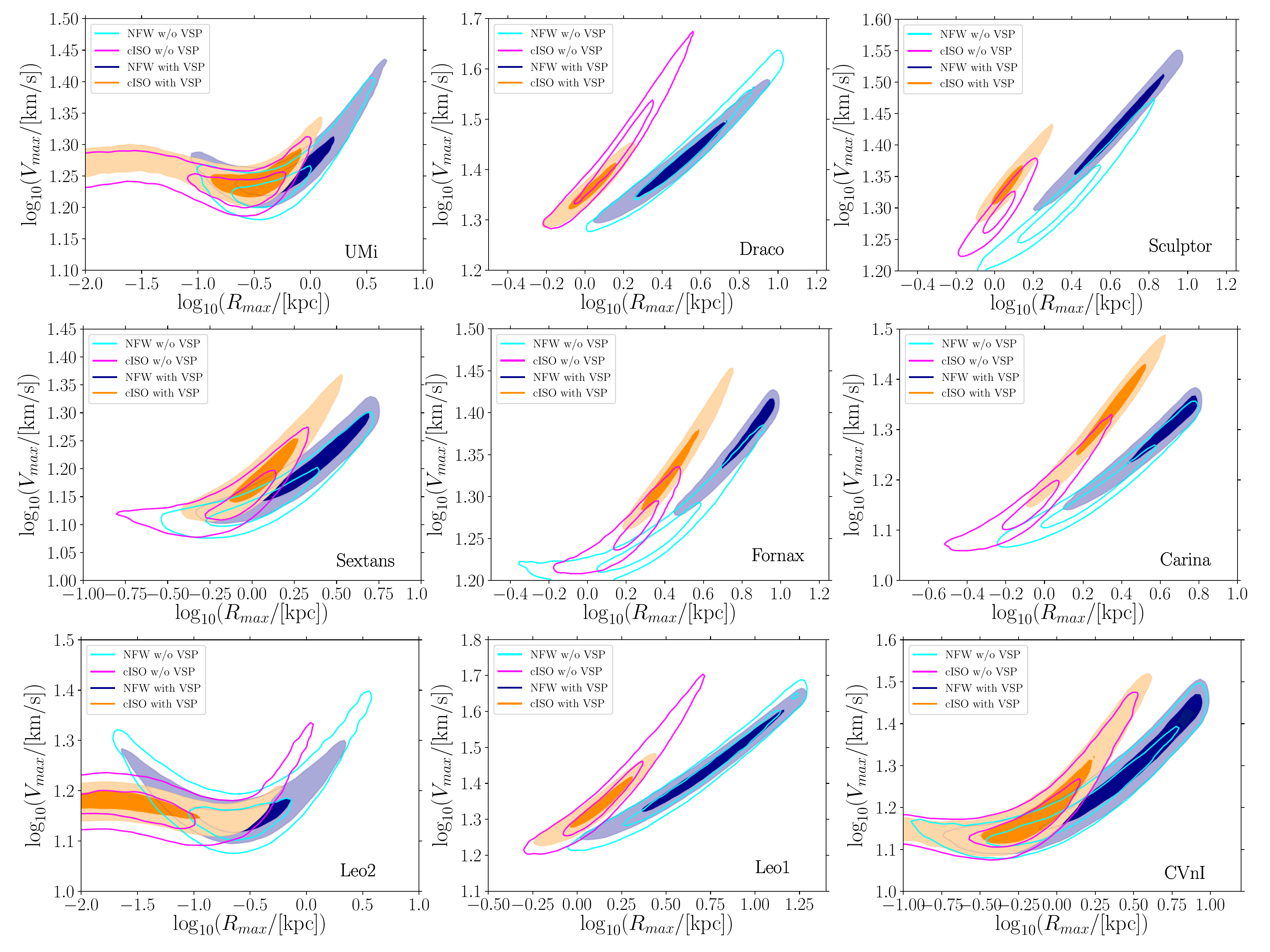}
}
\caption{\emph{68\% and 95\% highest probability density region in $\log_{10} \vmax$~-~$\log_{10} \rmax$ plane reported in figure~\ref{fig:MW_subhalo_diversity} for the bright MW dSphs, together with the results without including the constraint from the fourth-order virial shape parameter (VSP) $v_{s1}$.}}
\label{fig:fourth_order_estimator}
\end{figure}

Finally -- in order to highlight the effects of fourth-order projected virial theorems in our analysis -- we show in figure~\ref{fig:fourth_order_estimator} the 68\% and 95\% highest probability density region for $(\vmax,\rmax)$ with and without the information form the fourth-order virial shape parameters. The results for both cored and cuspy halo profiles are shown in figure~\ref{fig:fourth_order_estimator}.

\begin{table}
\centering
\renewcommand{\arraystretch}{1.5}
{\footnotesize
\begin{tabular}{|c|c|ccc|c|c|c|}
\hline
\textbf{MW dSph} & \boldmath$\widehat{\nu_{s1}}\ [10^3 \ \textrm{\bf km}^4\textrm{\bf /s}^4] $ & & \boldmath$\log_{10}(\vmax/[\textrm{\bf km/s}])$ & \boldmath$\log_{10}(\rmax/[\textrm{\bf kpc}])$ & \boldmath$\chi^2/\textrm{\bf d.o.f.}$ & \boldmath$\Delta \textrm{\bf AIC}$ & \boldmath$\Delta \textrm{\bf BIC}$ \\
\hline
\multirow{2}*{Ursa Minor (Umi)} & \multirow{2}*{$3.73^{+0.75}_{-0.62}$}
& \ NFW & $1.26^{+0.07}_{-0.03}$ & $-0.19^{+0.46}_{-0.46}$  & 0.95 & \multirow{2}*{0} &  \multirow{2}*{0} \\
 & & \ cISO & $1.24^{+0.03}_{-0.02}$ &  $-0.75^{+0.41}_{-0.77}$ & 0.94 & & \\
\hline
\multirow{2}*{Draco} & \multirow{2}*{$3.38^{+0.54}_{-0.46}$}
& \ NFW & $1.42^{+0.08}_{-0.07}$ & $0.50^{+0.23}_{-0.24}$  & 1.0 & \multirow{2}*{-3} &  \multirow{2}*{-3} \\
 & & \ cISO & $1.36^{+0.05}_{-0.04}$ &  $0.03^{+0.12}_{-0.13}$ & 0.99 & & \\
\hline
\multirow{2}*{Sculptor} & \multirow{2}*{$3.94^{+0.43}_{-0.37}$}
& \ NFW & $1.43^{+0.10}_{-0.08}$ & $0.62 ^{+0.29}_{-0.25}$  & 0.95 & \multirow{2}*{8} &  \multirow{2}*{4} \\
 & & \ cISO & $1.35^{+0.04}_{-0.03}$ &  $0.06^{+0.11}_{-0.08}$ & 0.99 & & \\
\hline
\multirow{2}*{Sextans} & \multirow{2}*{$1.84^{+0.37}_{-0.29}$}
& \ NFW & $1.19^{+0.07}_{-0.06}$ & $0.31^{+0.27}_{-0.34}$  & 0.95 & \multirow{2}*{-1} &  \multirow{2}*{-1} \\
 & & \ cISO & $1.18^{+0.07}_{-0.04}$ &  $0.05^{+0.21}_{-0.28}$ & 0.95 & & \\
\hline
\multirow{2}*{Fornax} & \multirow{2}*{$7.12^{+0.52}_{-0.48}$}
& \ NFW & {$1.36^{+0.04}_{-0.05}$} & {$0.80^{+0.12}_{-0.19}$}  & 0.96 & \multirow{2}*{10} &  \multirow{2}*{4} \\
 & & \ cISO & {$1.33^{+0.06}_{-0.04}$} &  {$0.45^{+0.17}_{-0.14}$} & 0.99 & & \\
\hline
\multirow{2}*{Carina} & \multirow{2}*{$1.79^{+0.33}_{-0.25 }$}
& \ NFW & $1.28^{+0.06}_{-0.08}$ & $0.57^{+0.16}_{-0.26}$  & 1.0 & \multirow{2}*{-1} &  \multirow{2}*{-2} \\
 & & \ cISO & $1.14^{+0.07}_{-0.05}$ &  $-0.09^{+0.18}_{-0.32}$ & 1.0 & & \\
\hline
\multirow{2}*{Leo~II} & \multirow{2}*{$1.46^{+0.41}_{-0.32}$}
& \ NFW & $1.16^{+0.06}_{-0.03}$ & $-0.58^{+0.47}_{-0.49}$  & 1.0 & \multirow{2}*{1} &  \multirow{2}*{2} \\
 & & \ cISO & $1.17^{+0.04}_{-0.03 }$ &  $-1.30^{+0.68  }_{-0.46 }$ & 1.0 & & \\
\hline
\multirow{2}*{Leo~I} & \multirow{2}*{$3.42^{+0.67}_{-0.56}$}
& \ NFW & {$1.42^{+0.13}_{-0.12}$} &  {$0.70 ^{+0.35}_{-0.38}$}  & 1.0 & \multirow{2}*{-2} &  \multirow{2}*{-2} \\
 & & \ cISO &  {$1.33^{+0.07}_{-0.06}$} &  {$0.12^{+0.17}_{-0.19}$} & 0.99 & & \\
\hline
\multirow{2}*{Canes Venatici~I (CVnI)} & \multirow{2}*{$3.13^{+1.33}_{-0.77}$}
& \ NFW & $1.24^{+0.15}_{-0.10}$ & $0.33^{+0.44 }_{-0.60}$  & 0.98 & \multirow{2}*{0} &  \multirow{2}*{-1} \\
 & & \ cISO & $1.19^{+0.15}_{-0.06}$ & $-0.09^{+0.38}_{-0.41}$ & 0.99 & & \\
\hline
\end{tabular}
}
\caption{\em Results from the Bayesian inference of the DM content in the nine bright MW dSphs with the refined Jeans analysis described in the text. From left to right we report in order: the name of the dSph, the data-based estimator for the fourth-order virial shape parameter adopted in our analysis (16-$th$,50-$th$,84-$th$ percentile), the inferred $\vmax$ and $\rmax$ pair (16-$th$,50-$th$,84-$th$ percentile), $\chi^2$/d.o.f. related to equation~\eqref{eq:Lvlos} (yielding a rough estimate on the goodness of the residuals in the fit to stellar line-of-sight velocities), and the difference in the Akaike Information Criterion and Bayesian deviance Information Criterion (see text) between NFW and cISO scenarios. Positive differences of more than few units~\citep{BayesFactors} a preference for cISO and negative ones point to a preference for the NFW profile.
\label{tab:results}}
\end{table}

\section{Too-big-to-fail for cuspy and cored halo profiles}
\label{app:B}
\begin{figure}
\vspace{10mm}
\centering{
\includegraphics[scale = 0.6]{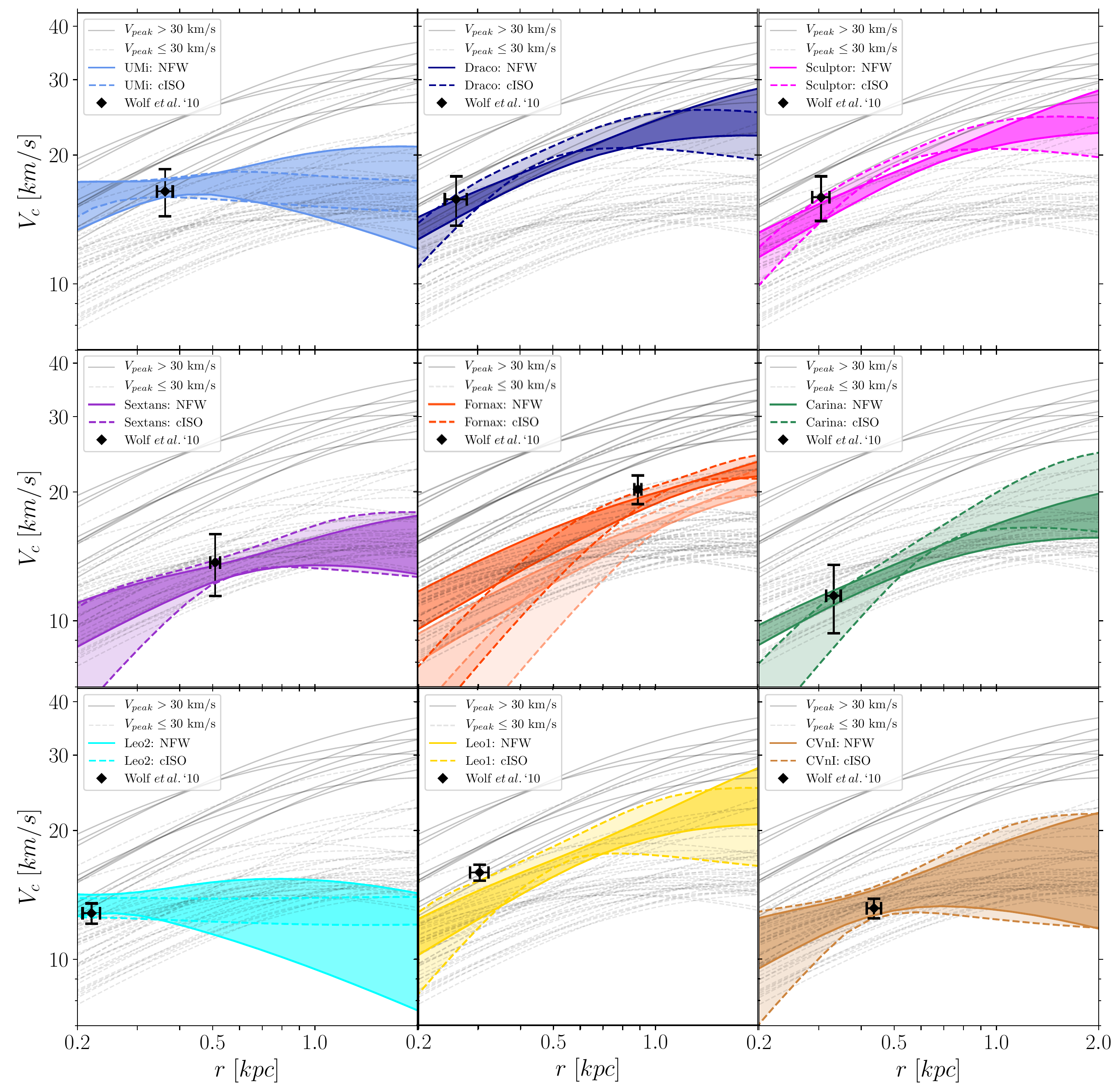}
}
\caption{\emph{Too-big-to-fail problem of the bright MW dSphs for both cuspy and cored scenarios. Colored bands encompass the 68\% highest probability density region and correspond to the posterior distribution of the circular velocity profile {related to the DM halo of the dSph} computed at different radii. 
{For the case of Fornax we also show the total circular velocity profile, comprising the stellar contribution of this galaxy}.
Light-gray band represents the prediction for the circular velocity profile of the 50 most massive subhalos (10 most massive featuring $V_{\rm peak}>30$ km s$^{-1}$) in the ELVIS simulation Kauket of~\citet{Garrison-Kimmel:2013eoa}. Data points in black corresponds to the mass estimator of~\citet{Wolf:2009tu}, updated with the photometric and spectroscopic information collected recently in~\citet{MNRAS:Simon2019}.}}
\label{fig:TBTF_classical}
\end{figure}

We have seen how cuspy and cored DM halos in our detailed investigation lead to the very same quality of the fit on dSph stellar kinematics in the absence of any cosmological prior imposed on ($\vmax$,$\rmax$) pair. Here we reinforce the picture already drawn in the main body of this manuscript about how both cuspy and cored DM scenarios are both subject to the too-big-to-fail problem. We present in figure~\ref{fig:TBTF_classical} the $68\%$ region of the posterior distribution of the circular velocity profile obtained for both cuspy and cored fits. In the same figure, we show the band encompassing the prediction for the circular velocity profiles of the most massive subhalos ($V_{peak}>30$ km/s) from the CDM ELVIS simulation Kauket~\citep{Garrison-Kimmel:2013eoa}. We also present the mass estimator data point $V_{1/2}$ at the deprojected half-light radius~\citep{Wolf:2009tu}. In order to compare with the literature in a quite agnostic manner, we update the values reported in~\citet{Wolf:2009tu} using the average line-of-sight velocity dispersions collected in~\citet{MNRAS:Simon2019} (our own $\overline{\sigma}^2_{los}$ results compatible within errors) and, most importantly, including the new photometric dataset from \citep{MNRAS:Munoz2018}. 
Figure~\ref{fig:TBTF_classical} shows the tension between the CDM N-body prediction and the outcome of our fits with cored and cuspy DM halos for the classical dSphs. It also shows that the mass estimator works well for both profiles. This is a non-trivial check since the mass estimator was derived with only information from the dispersion and not the fourth-order projected virial theorem.

\section{Details and cross-checks on the \boldmath$\rho_{150}\textup{--}r_{\rm P}$ correlation}
\label{app:C}
We provide here further details and validation of the correlation discussed in the main text for the nine bright MW dSphs, involving a relation between DM densities at 150 pc, $\rho_{150}$, and their pericenter distance, $r_{\rm P}$, estimated from the Gaia data~\citep{Fritz:2018aap}. To formally establish such correlation, we perform a power-law fit for the nine bright MW dSphs based on the basis of the following test statistic:
\begin{eqnarray}
    \label{eq:rho_pericenter}
    \mathcal{L}_{\textrm{corr}} = \prod_{i=1}^{9} \frac{\exp\left(-\frac{1}{2} \chi^2_{\textrm{corr},i}\right)}{\sqrt{2 \pi \left(\Delta x_{i}^2 + \delta y_{i}^2\right)}} \ \ , \ \
    \chi^2_{\textrm{corr},i} = \frac{\left(y_{i} - 10^q \, x_{i}^m \right)^2}{\Delta x_{i}^2 + \delta y_{i}^2} \ \ , \ \ \Delta x_{i} = m \, 10^q x_{i}^{m-1} \delta x_{i} \ ,
\end{eqnarray}
where for each of the objects: $x = r_{\rm P}/\textrm{kpc}$, $y = \rho_{150}/(10^7M_{\odot}\textrm{kpc}^{-3})$ and $\delta x$, $\delta y$ are the estimated errors. In order to infer $\rho_{150}$ and $\delta \rho_{150}$ we compute the 16-$th$, 50-$th$ and 84-$th$ percentile of the posterior distribution of the DM density at 150 pc, inferred from the Bayesian analysis of classical dSphs described in detail in appendix~\ref{app:A}.
The quoted values for $r_{\rm P}$ and $\delta r_{\rm P}$ are taken from~\citet{Fritz:2018aap}, who provide this information for both a low-mass and high-mass MW model. 
For both density and pericenter distance we symmetrize errors and shift median values accordingly. We perform a MCMC analysis using again the affine-invariant sampling algorithm of~\citet{goodman2010}, implemented in the package \textit{emcee}~\citep{ForemanMackey:2012ig}.
\begin{figure}
\vspace{10mm}
\centering{
\includegraphics[scale = 0.47]{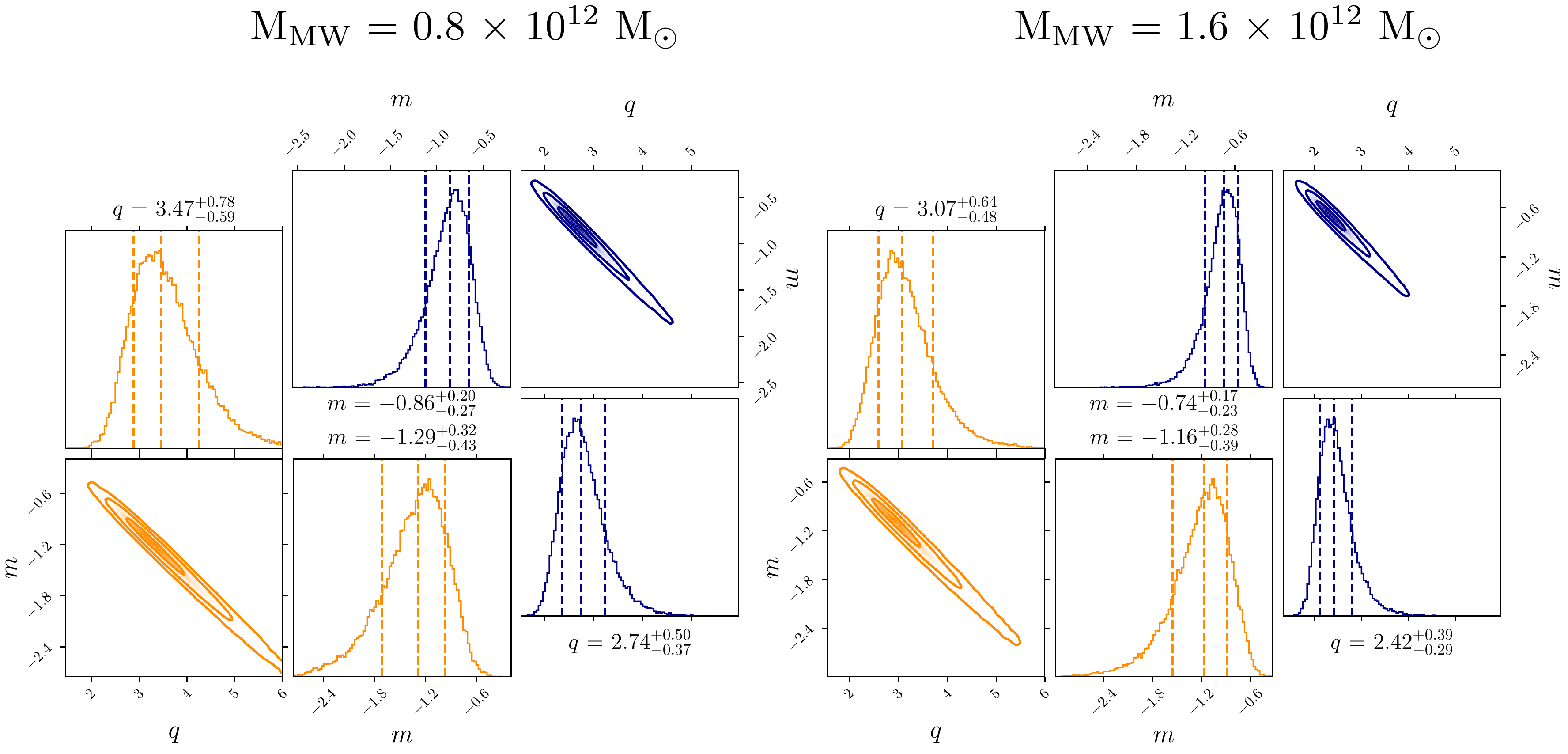}
}
\caption{\emph{Posterior probability density functions for the parameters defining the power-law relation between MW classical central densities, $\rho_{150}$, and pericenter distances, $r_{\rm P},$ see the likelihood given in equation~\eqref{eq:rho_pericenter}. We report in blue the result obtained for the cuspy DM density profile considered in this work, in orange the one for the scenario with DM inner-core Density. On the left (right) panel we show the outcome of our inference for the case where the estimate for dSph pericenters come from the assumption of MW mass equals to $0.8\,$($1.6$)~$10^{12}$ ${\rm M_{\odot}}$.}}
\label{fig:triangle_corr}
\end{figure}

In figure~\ref{fig:triangle_corr} we show the outcome of our inference on $q$ and $m$ parameters appearing in equation~\eqref{eq:rho_pericenter}. 
For both NFW and cISO cases, there is a tight $\rho_{150}$~-~$r_{\rm P}$ anti-correlation: no correlation case is well beyond the 99.99$\%$ highest probability density region (the largest contour drawn for the joint posterior probability distribution in the same figure).  
In both the low-mass and high-mass MW cases, we observe that the analysis with the NFW profile yields a more precise determination of two parameters.
We also show the results of adopting the pericenter distances corresponding to the high-mass MW model from~\citet{Fritz:2018aap}, as a point of comparison for the results in the main text. The same information that is in  figures~\ref{fig:corr_rho_pericenter}~-~\ref{fig:corr_rho_pericenter_UDGs} is plotted in the left and central panel of figure~\ref{fig:triangle_corr} for the MW model with mass $1.6 \times 10^{12}$~\textrm{M}$_{\odot}$. We note the inference of this correlation does not {critically}  depend on the assumed MW model. 

\begin{figure}
\vspace{10mm}
\centering{
\includegraphics[scale = 0.59]{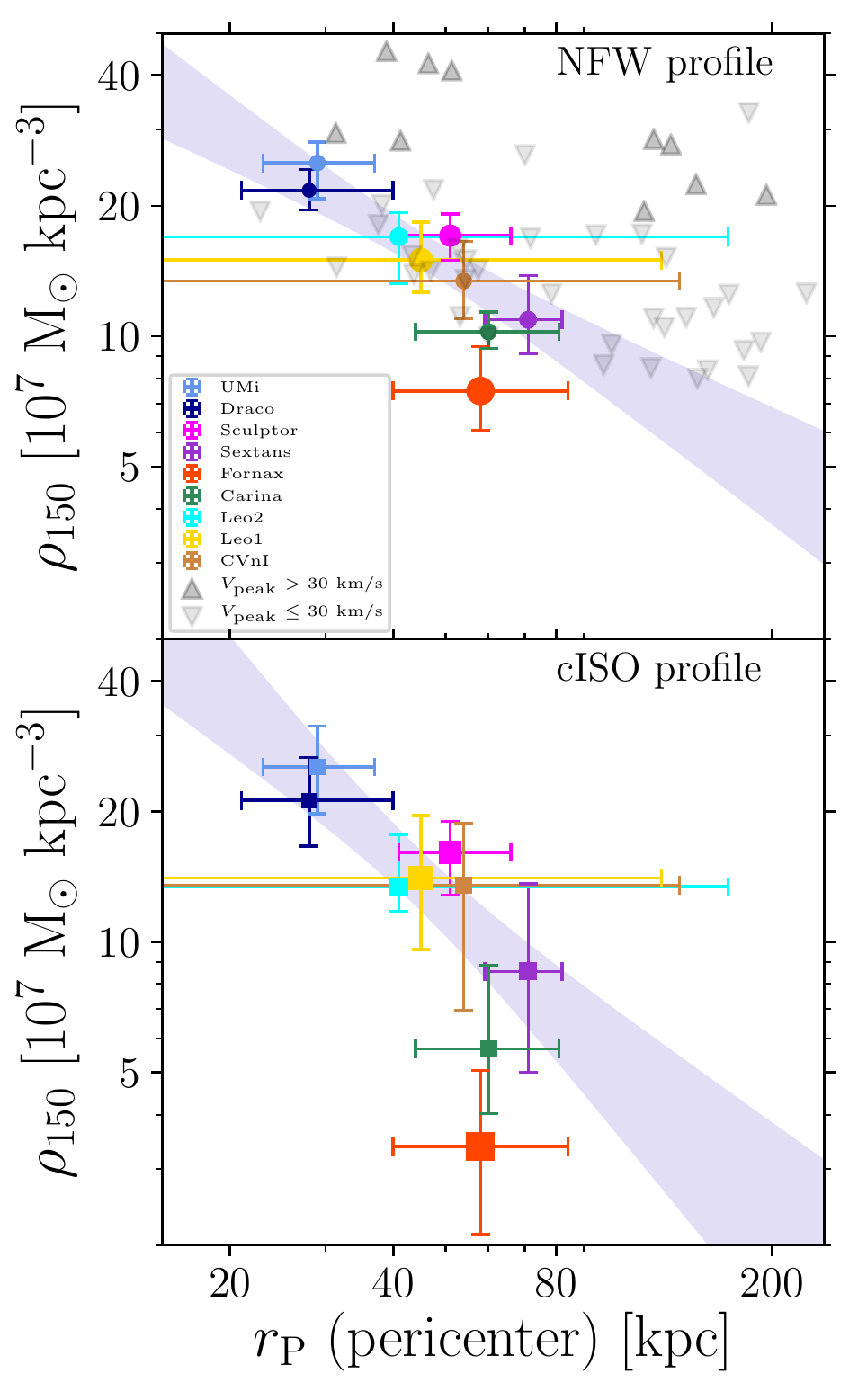}
\includegraphics[scale = 0.59]{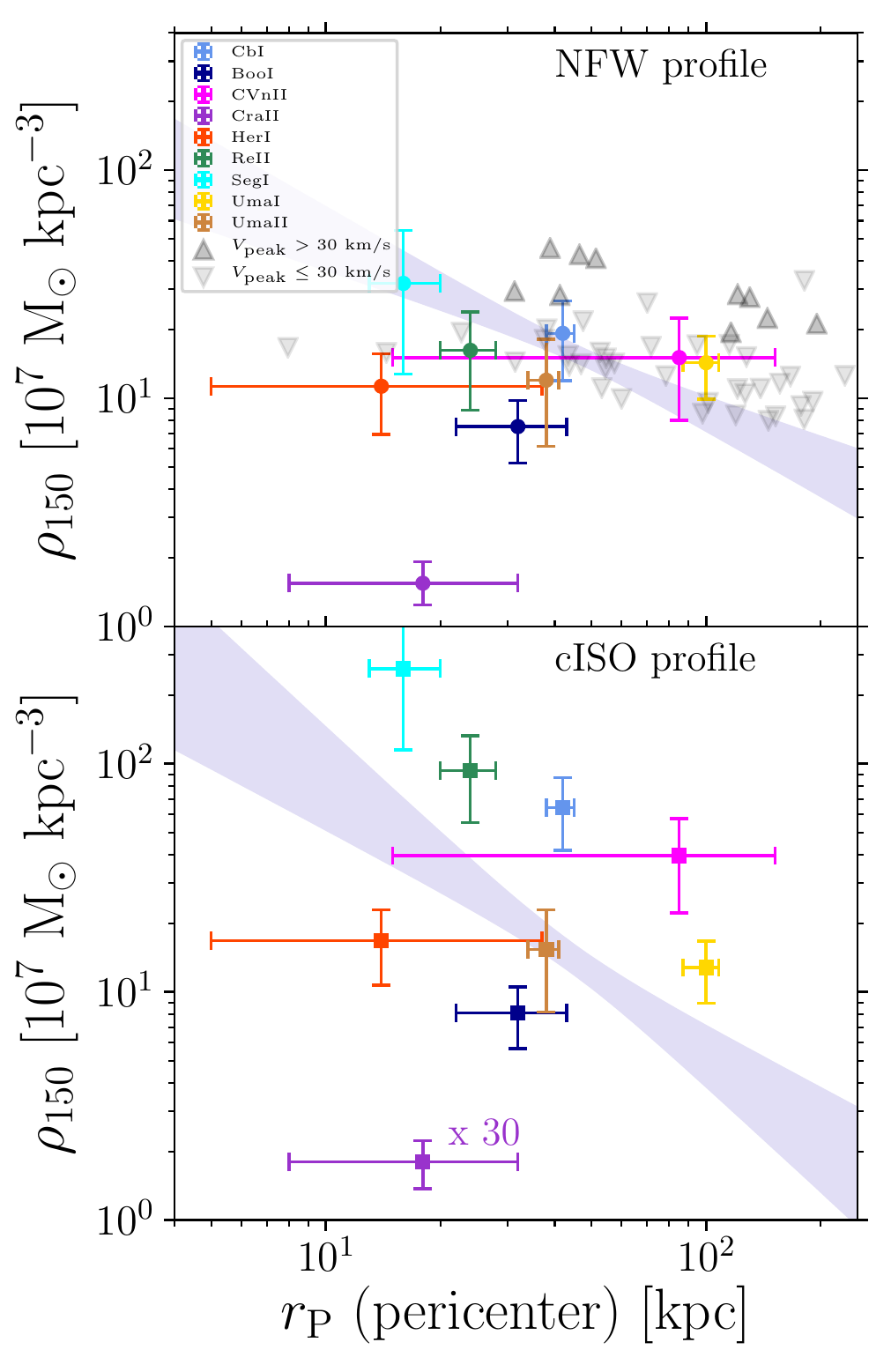}
\includegraphics[scale = 0.59]{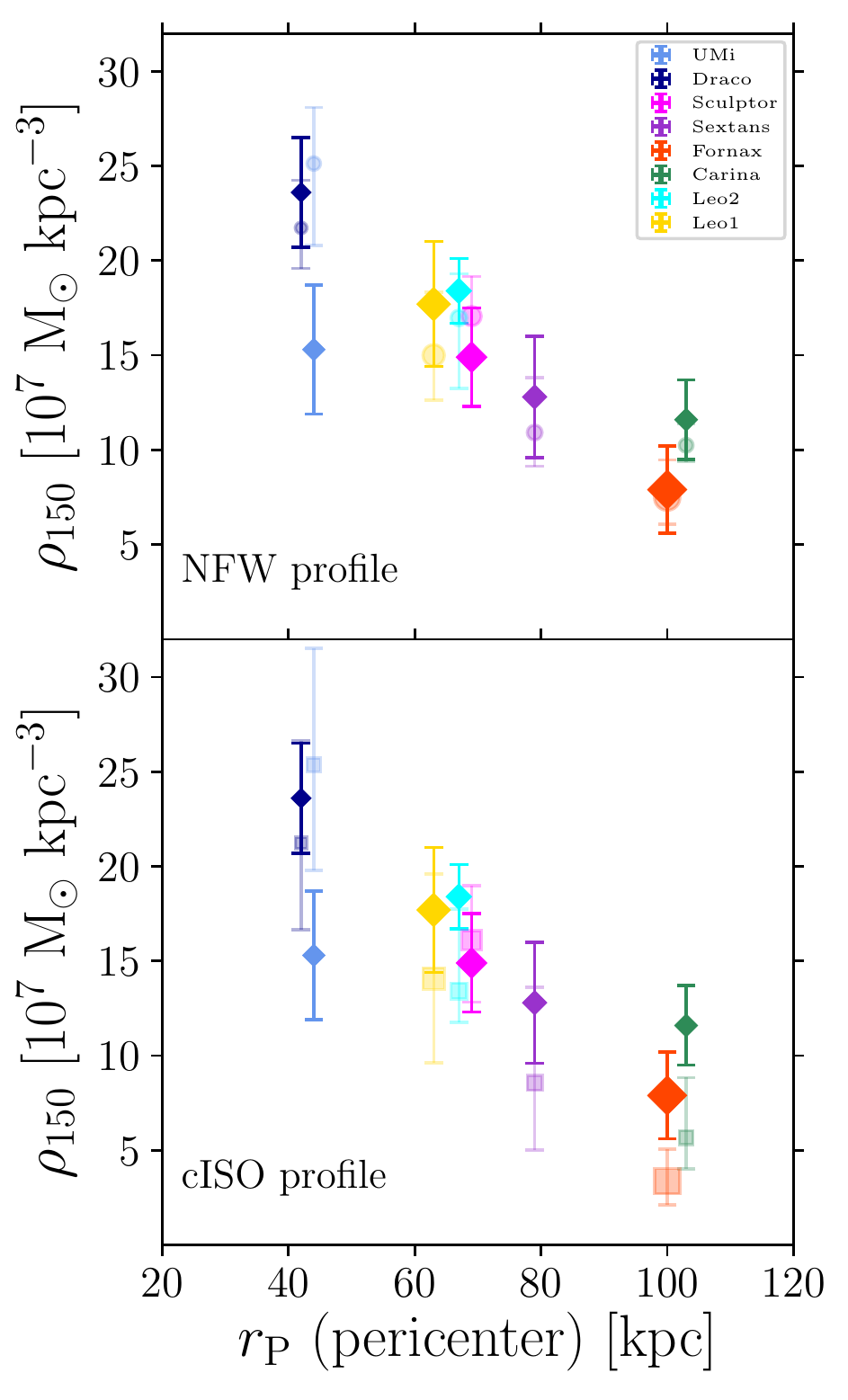}
}
\caption{\emph{Same as figure~\ref{fig:corr_rho_pericenter} (left panel) and figure~\ref{fig:corr_rho_pericenter_UDGs} (central panel) using the pericenter distance from the MW, $r_{\rm P}$, as estimated in~\citet{Fritz:2018aap} adopting a MW mass of $1.6 \times 10^{12}$~M$_{\odot}$. In the right panel, we show the direct comparison of our Bayesian inference of $\rho_{150}$ for the eight classical MW dSphs (shaded points) against $\rho_{150}$ quoted in~\citet{Read:2018fxs}, reported with diamond points.}}
\label{fig:corr_rho_pericenter_appendix}
\end{figure}

In the same figure we provide a direct comparison of the inferred densities at 150 pc with the ones obtained in~\citet{Read:2018fxs}. 
CVn~I has not been included in that study and therefore we restrict to the other eight bright dSphs. Figure~\ref{fig:corr_rho_pericenter_appendix} shows that the inferred $\rho_{150}$ for the classical dSphs in this work are in good agreement with the inferences of~\citet{Read:2018fxs}.

\begin{figure}
\vspace{10mm}
\centering{
\includegraphics[scale = 0.63]{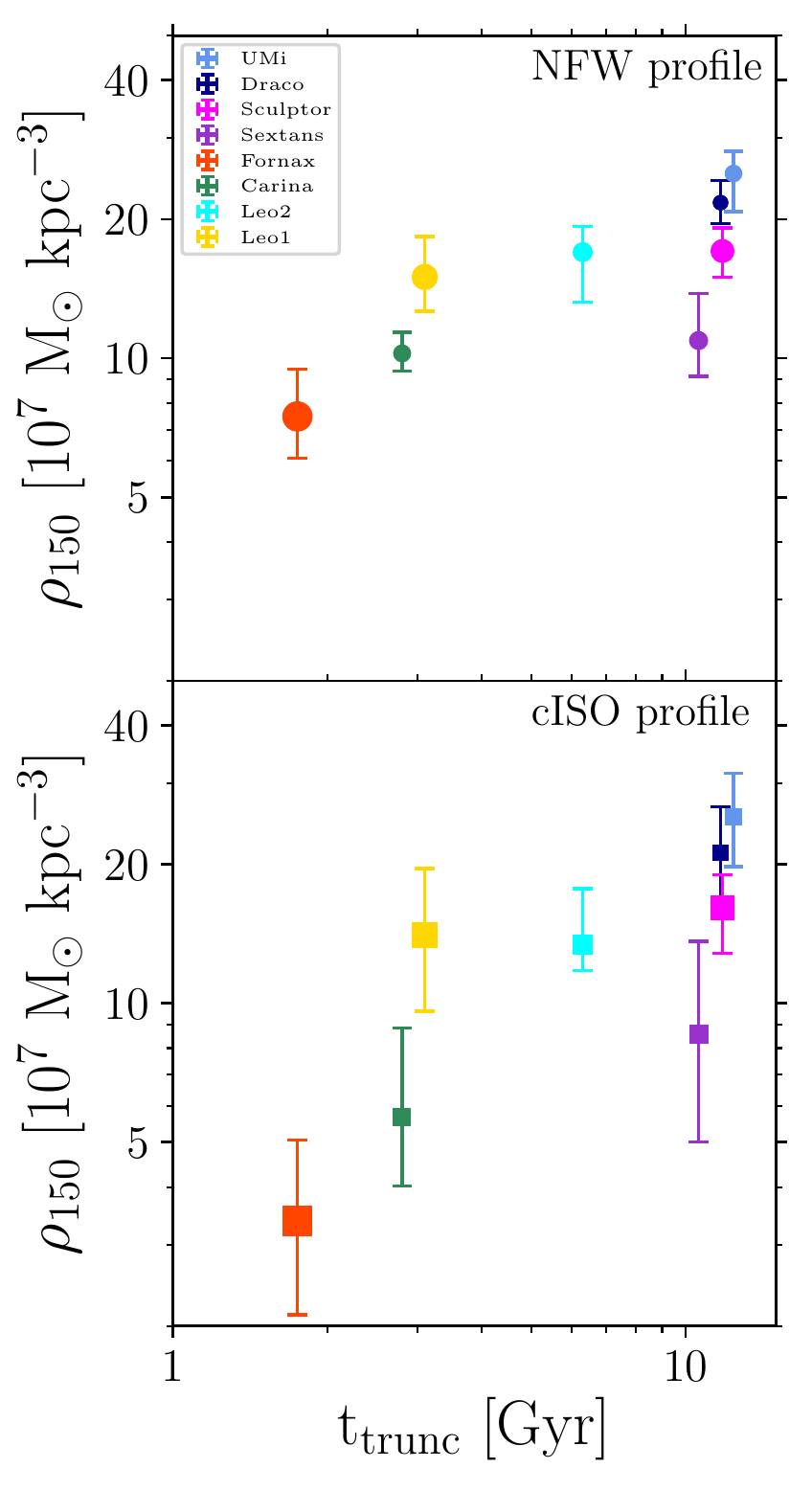}
\includegraphics[scale = 0.63]{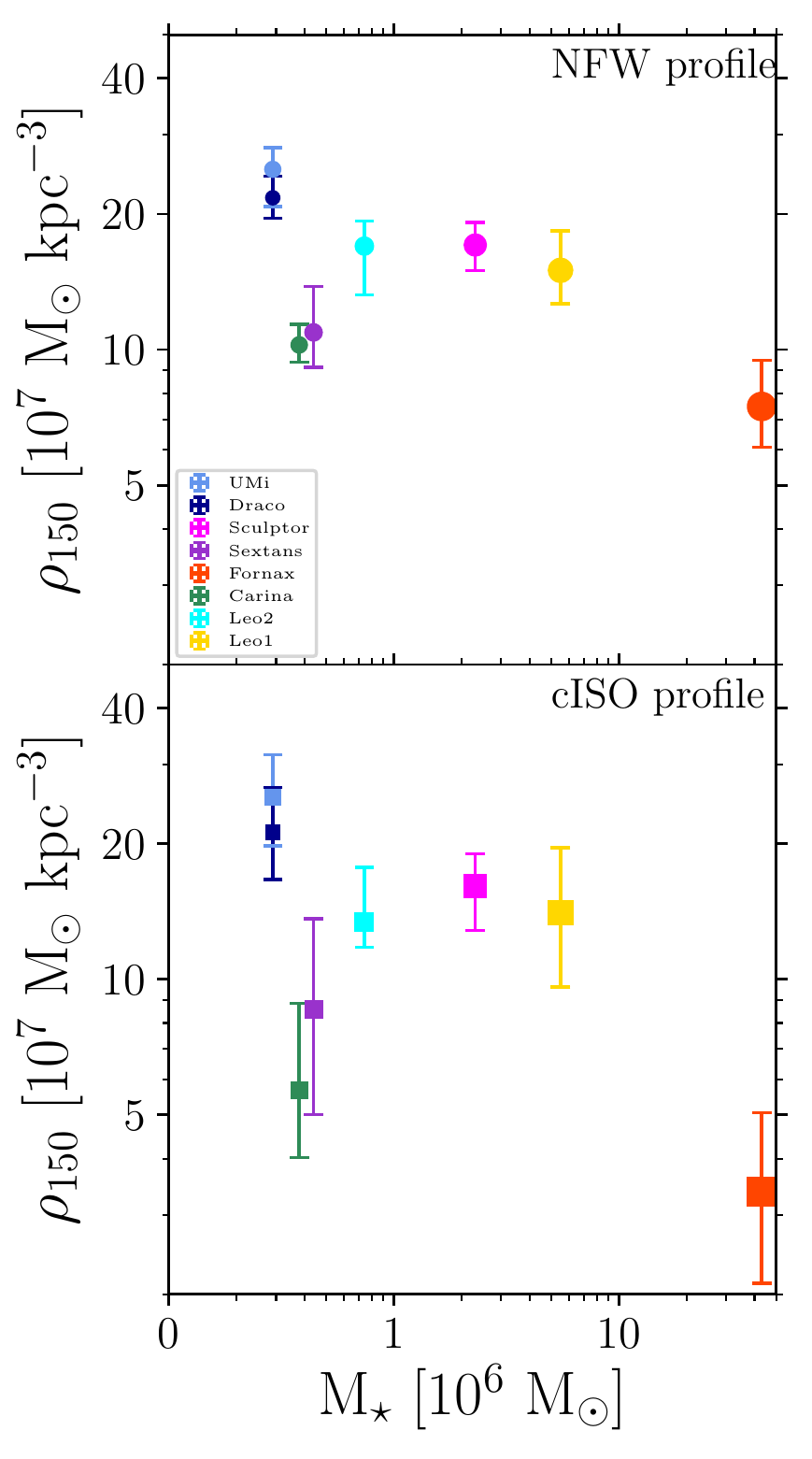}
\includegraphics[scale = 0.63]{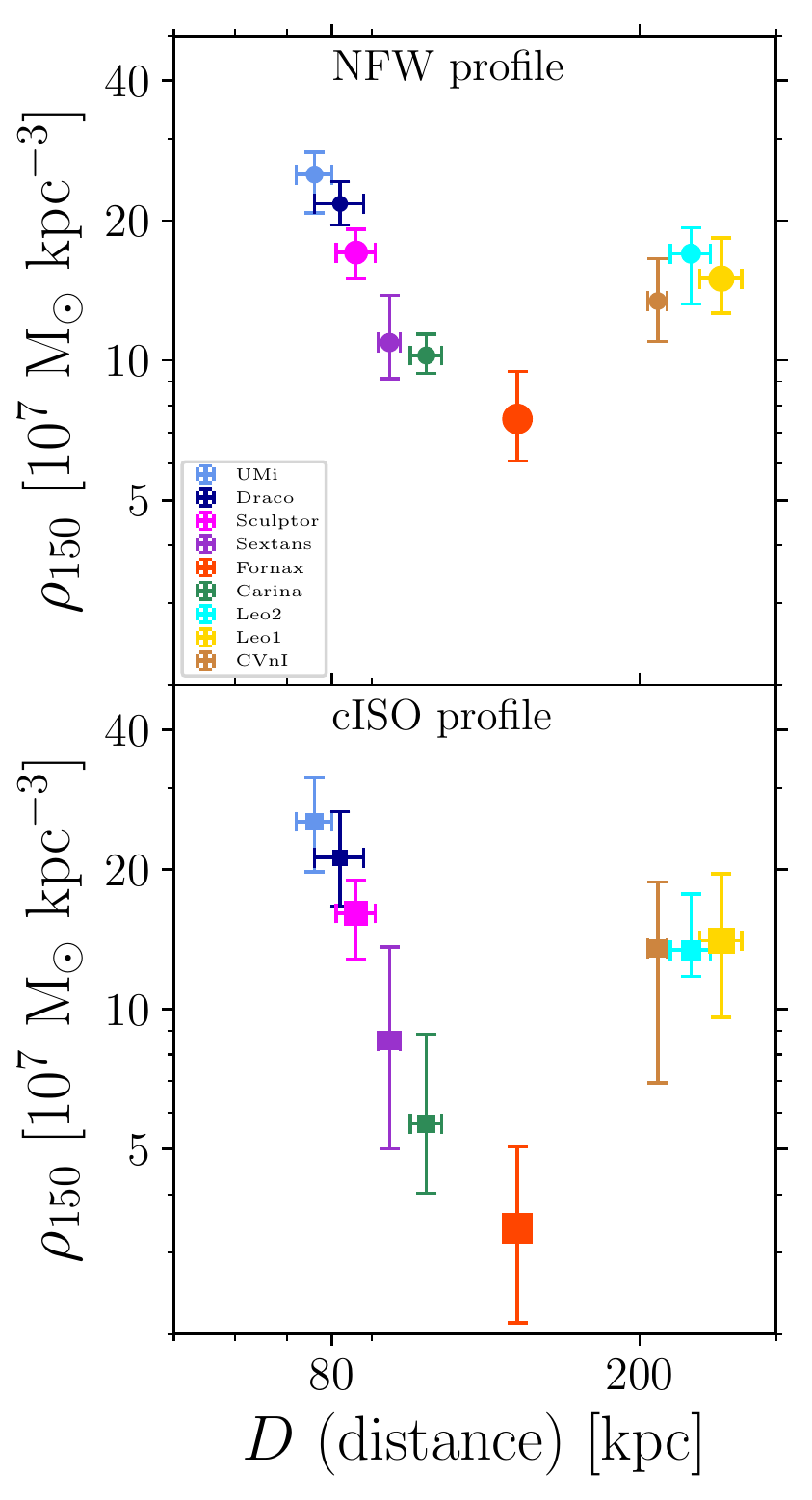}
}
\caption{\emph{Same as figure~\ref{fig:corr_rho_pericenter} (left panel) trading the pericenter distance from the MW, $r_{\rm P}$, for: star formation shutoff time as quoted in~\citet{Read:2018fxs}; dSph stellar mass as reported in the same reference; heliocentric distance from ~\citet{MNRAS:Munoz2018}.}}
\label{fig:further_tests_appendix}
\end{figure}

Finally, in figure~\ref{fig:further_tests_appendix} we provide further tests of the correlation inferred for the nine bright dSphs. In the left panel of the figure, we show our $\rho_{150}$ estimate versus the star formation shutoff time presented in~\citet{Read:2018fxs}; in the central panel we inspect the possible correlation of dSph central densities with their stellar mass counterpart; in the right panel of figure~\ref{fig:further_tests_appendix} we replace the pericenter distance with the heliocentric distance. While the first two panels may be informative with respect to what investigated in~\citet{Read:2018pft}, the last one may be useful in light of the recent claims provided in~\citet{Hammer:2018arx}. As mentioned in the main text, we highlight that no {strong} correlation emerges from the comparison of the central density of the bright dSphs against these three different quantities.

\bsp	
\label{lastpage}
\end{document}